\documentclass[apj,numberedappendix]{emulateapj}
\usepackage{natbib}
\def\xmms{{\it XMM-Newton} }
\def\suzakus{{\it Suzaku} }
\def\swifts{{\it Swift} }
\def\saxs{{\it Beppo-SAX} }
\def\rxtes{{\it RXTE} }
\def\integrals{{\it INTEGRAL} }
\def\kws{KW }
\def\xmm{{\it XMM-Newton}}
\def\suzaku{{\it Suzaku}}

\def\sax{{\it Beppo-SAX}}
\def\rxte{{\it RXTE}}
\def\integral{{\it INTEGRAL}}
\def\kw{KW}

\shorttitle{Diffuse Nonthermal Emission in the Coma Cluster}
\shortauthors{Wik et al.}

\begin{document}

\title{The Lack of Diffuse, Nonthermal Hard X-ray Emission 
in the Coma Cluster: \\ The \swifts BAT's Eye View}

\author{Daniel R. Wik\altaffilmark{1, 8}, Craig L. Sarazin\altaffilmark{2},
Alexis Finoguenov\altaffilmark{3,4}, Wayne H. Baumgartner\altaffilmark{1},
Richard F. Mushotzky\altaffilmark{6}, Takashi Okajima\altaffilmark{1},
Jack Tueller\altaffilmark{1}, and Tracy E. Clarke\altaffilmark{7}}

\altaffiltext{1}{Astrophysics Science Division, Laboratory for High 
Energy Astrophysics, Code 662, NASA/Goddard Space Flight Center,
Greenbelt, MD 20771, USA; daniel.r.wik@nasa.gov}
\altaffiltext{2}{Department of Astronomy, University of Virginia}
\altaffiltext{3}{Max Planck Institute for Extraterrestrial Physics}
\altaffiltext{4}{Center for Space Science Technology, University of 
Maryland-Baltimore County}
\altaffiltext{5}{Joint Center for Astrophysics, University of 
Maryland-Baltimore County}
\altaffiltext{6}{Department of Astronomy, University of Maryland, 
College Park}
\altaffiltext{7}{Naval Research Laboratory}
\altaffiltext{8}{NASA Postdoctoral Program Fellow}

\begin{abstract}
The Coma cluster of galaxies hosts the brightest radio halo known
and has therefore been the target of numerous searches for associated
inverse Compton (IC) emission, particularly at hard X-ray energies
where the IC signal must eventually dominate over thermal emission.
The most recent search with the \suzakus Hard X-ray Detector (HXD)
failed to confirm previous IC detections 
with \rxtes and \sax, instead setting an upper limit 
2.5 times below their nonthermal flux.
However, 
this discrepancy can be resolved if the IC emission is
very extended, beyond the scale of the cluster radio halo.
Using reconstructed sky images from the 58--month \swifts BAT 
all sky survey, the feasibility of such a solution is investigated.
Building on Renaud et al., we test and implement a method for extracting
the fluxes of extended sources, assuming specified spatial distributions.
BAT spectra are jointly fit with an \xmms EPIC-pn spectrum
derived from mosaic observations.
We find no evidence for large-scale IC emission 
at the level expected from
the previously detected nonthermal fluxes.
For all nonthermal spatial distributions considered, which span the gamut
of physically reasonable IC models, we determine upper limits
for which the largest (most conservative) limit is
$\la 4.2\times10^{-12}$ erg s$^{-1}$ cm$^{-2}$ (20--80 keV), 
which corresponds to a
lower limit on the magnetic field $B > 0.2$ $\mu$G.
A nominal flux upper limit of $< 2.7\times10^{-12}$ 
erg s$^{-1}$ cm$^{-2}$, with
corresponding $B > 0.25$ $ \mu$G, is derived for the most probable IC
distribution given the size of the radio halo and likely magnetic
field radial profile.
\end{abstract}


%------------------------------ INTRO ------------------------------
\section{Introduction}
\label{sec:comabat:intro}

The X-ray emission from clusters of galaxies is primarily thermal
in origin and is produced by a diffuse population of intergalactic
electrons in the ionized intracluster medium (ICM).
These electrons coexist with a nonthermal,
relativistic electron population in at least some clusters -- inferred 
from observations in the radio regime -- which should also radiate at
X-ray energies.
While thermal emission clearly dominates in the kilo-electron volt (keV)
energy range, it declines rapidly outside this range, allowing the
detection of a nonthermal spectral signature as soft or hard excess
emission.
This possibility is especially promising at hard ($>$10 keV) energies,
where the exponential decline of the thermal bremsstrahlung continuum
is distinctly steeper than the expected nonthermal spectrum.
Measurements of nonthermal X-ray emission are critical to
the determination of the total amount of relativistic energy
in the ICM, which is currently poorly constrained.
While no more than $\sim$10\% of this energy is tied up in
nonthermal components, amounts at or near this level will affect the
dynamics and structure of the thermal gas \citep[e.g.,][]{VBG09}.
Specifically, studies that attempt to infer the total masses of
clusters from the hydrostatic state of the thermal gas will produce
biased mass estimates if the pressure support of relativistic
particles and fields is not accurately included.
The mass functions built from these estimates
can be used to constrain cosmological
parameters; these studies are already underway using observables
derived in both the X-ray \citep[e.g.,][]{MAE+08, Vik+09} and 
microwave \citep[e.g.,][through the Sunyaev-Zel'dovich effect]{Van+10} 
regimes.

A measurement of the total energy in relativistic ICM components is
possible when X-ray and radio nonthermal fluxes are combined.
Diffuse, cluster-wide synchrotron radio emission, called radio halos or
relics depending on their morphology, imply that both magnetic fields and
relativistic electron populations are present on large scales.
The total luminosity of a synchrotron-emitting electron is given by
\begin{equation} \label{eq:lsync}
L_R = \frac{4}{3}\sigma_Tc\gamma^2\epsilon_{B}
\, ,
\end{equation}
where $\sigma_T$ is the Thomson cross-section, $c$ is the speed of light,
$\gamma$ is the Lorentz factor of the electron, and 
$\epsilon_B=B^2/8\pi$ is the energy density of the magnetic field.
For a collection of relativistic electrons, 
the value of $L_R$ depends both on the number of
electrons and on $B$ and cannot independently determine either.
However, these same electrons will up-scatter cosmic microwave
background (CMB) photons through inverse Compton (IC) interactions, which have
a luminosity $L_X$ equivalent in form to equation~(\ref{eq:lsync}) but with
$\epsilon_B$ replaced by the energy density of the CMB.
Since both luminosities are proportional to the number of electrons, 
their ratio gives the volume-averaged magnetic field,
\begin{equation} \label{eq:comabat:synicratio}
\frac{L_R}{L_X} = \frac{B^2/8\pi}{aT_{CMB}^4}
\, ,
\end{equation}
where $a$ is the radiation constant and $T_{CMB}$ is the temperature of the
CMB.
The IC radiation should be observable at hard X-ray energies \citep{Rep77}.
Thus far, IC emission has only been detected at low significance
\citep{NOB+04, MA09} or, in one case at higher significance, 
in the Ophiuchus cluster (\citealt{EPP+08}; but see also 
\citealt{Aje+09} and \citealt{Fuj+08}),
although the diffuse radio emission in this cluster is restricted to a 
smaller scale mini-halo \citep{Mur+10}.
The measurement of an IC flux from a synchrotron source directly leads
to a simultaneous determination of the average value of $B$ and
the relativistic electron density \citep{HR74, Sar88}.
Therefore searches for IC emission coincident with a radio
halo or relic are an excellent way to constrain the
contribution of relativistic materials in clusters.

The first, and brightest, radio halo was discovered by \citet{Wil70} in
the Coma cluster, and its radio properties have perhaps been the best
studied \citep[e.g.][]{GFV+93, DRL+97, TKW03}.
Coma has been observed by all the major observatories with hard X-ray
capabilities \citep{HM86, Baz+90, HBS+93, RUG94},
and more recently non-thermal detections have been
claimed by \citet{RG02} with {\it RXTE} and by
\citet{FDF+99, FOB+04} with {\it BeppoSAX}, though
the latter detection is controversial \citep{RM04, FLO07}.
Due to the large field of view (FOV) of these non-imaging instruments
and the simple characterization of the thermal gas, the source
of this emission remains uncertain.
Even more recently, long ($\sim 1$ Msec) observations with {\it INTEGRAL}
have imaged extended diffuse hard X-ray emission from Coma, though it
was found to be completely consistent with thermal emission
\citep{RBP+06, ENC+07, LVC+08}.

Most recently, \citet{WSF+09} performed a joint analysis of spectra 
from the \xmms EPIC-pn and \suzakus HXD-PIN instruments -- the most 
sensitive instruments at soft and hard energies to date -- of the Coma 
cluster and were unable to detect IC emission.
Instead, they found an upper limit 2.5 times below the detections of 
\citet{RG02} and \citet{FOB+04}.
However, the narrower FOV of the HXD relative to the
collimators of the \rxtes PCA/HEXTE and \saxs PDS leaves open the
possibility that the spatial distribution of IC photons is
highly
extended, and therefore much of the flux was missed by the HXD.
The IC would have to be much broader than the size of the radio halo.
A uniform IC surface brightness of at least 30\arcmin\ in radius 
from the cluster center is sufficient to reconcile these results.
Therefore, an imaging analysis at hard X-rays is required to
confirm this picture;
unfortunately, no focussing hard ($>$10 keV) X-ray telescope 
has yet been deployed.
In the meantime, it is possible to perform a crude imaging analysis
with coded mask instruments, as previously discussed by \citet{RGL+06}.

In this work, we report on the spatial and spectral hard X-ray emission
from the Coma cluster using the 58--month accumulation of the \swifts 
Burst Alert Telescope (BAT) all-sky survey.
After the first 9 months of the survey, Coma was seen to be
clearly extended \citep{Aje+09}, so an accurate measurement of its 
flux must account for its resolved nature;
the standard method of extracting fluxes from coded mask instruments
assumes the underlying source to be point-like.
Using models for the spatial distribution of thermal and potential
nonthermal emission, we measure the total, extended flux in the 8
energy bands that make up the survey.
These fluxes are then converted into spectra, which we jointly fit
with an \xmms EPIC-pn spectrum from a {\it spatially identical
region}.
In this way, despite poor spatial resolution 
($\sim20$\arcmin), we are sensitive to any large-scale, extended
emission above the detection threshold for the survey.
While the sensitivity of the BAT detector is lower than instruments
such as the \suzakus HXD-PIN, the survey's large exposure time
-- thanks to a FOV that sees $1/8^{\rm th}$ of the sky
in a single pointing -- gives it a comparable, if not superior,
overall sensitivity.
In Section~\ref{sec:comabat:obs}, we describe the \swifts BAT survey 
in general and the \xmms EPIC-pn and BAT observations of the Coma
cluster specifically.
The extraction of spatially extended fluxes from models, along with
the specific models themselves, is discussed in 
Section~\ref{sec:comabat:spatial}.
Spectra constructed from these spatial fits are presented in
Section~\ref{sec:comabat:specfits}, along with the results of
joint fits with the \xmms spectrum.
In Section~\ref{sec:comabat:uls}, we provide upper limits on
spatially extended, nonthermal emission,
and in Section~\ref{sec:comabat:disc} we discuss the implications
of our non-detection for the relativistic phase of the ICM of the
Coma cluster.
In the appendices, we describe the calibration of the survey such
that joint fits with \xmms are straightforward, and we demonstrate
that the BAT instrument intrinsically detects extended emission on
the scales of interest here, though with higher uncertainty
than for a point source.
We assume a flat cosmology with $\Omega_M = 0.23$ and $H_0 = 72$ km/s/Mpc
and a luminosity distance to Coma of 98.4 Mpc.
Unless otherwise stated, all uncertainties are given at the 90\% 
confidence level.

%------------------------------ OBS --------------------------------
\section{Observations}
\label{sec:comabat:obs}

To achieve the necessary spatial coverage and spectral sensitivity,
we take advantage of mosaics of the Coma cluster constructed from
observations by the \xmms and \swifts satellites.
The high sensitivity and good spectral and spatial resolution of the
\xmms EPIC-pn data act as a check on the interpretation of the \swifts BAT
data, allowing the thermal and potentially nonthermal emission at
hard energies to be accurately decoupled.

\subsection{\xmms EPIC-pn Mosaic Observations}
\label{sec:comabat:obs:xmm}

\begin{figure}
\plotone{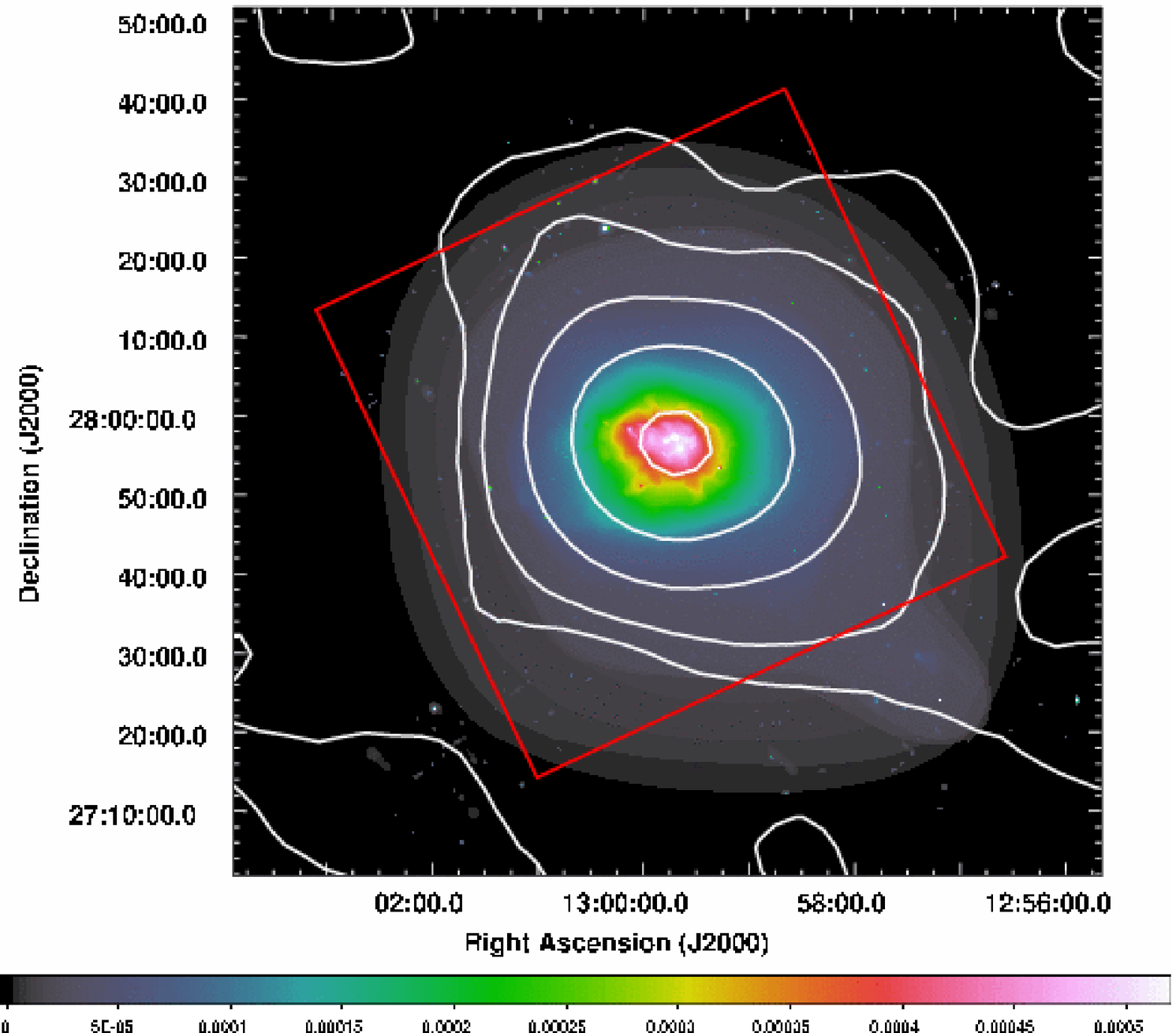}
\caption{\xmms EPIC-pn 2--7.5 keV wavelet-smoothed X-ray surface
brightness image \citep{SFM+04} with contours from the raw
\swifts BAT 14--20 keV survey image (square root spacing: $0.0$,
$2.625\times10^{-6}$, $1.05\times10^{-5}$, $2.3625\times10^{-5}$, 
and $4.2\times10^{-5}$
counts s$^{-1}$ pix$^{-1}$).
Negative contours are not shown for clarity, and note that the
FWHM of the BAT PSF is $19\farcm5$.
The (red) box shows the $65\farcm5\times65\farcm5$ region from which
the EPIC-pn spectrum is extracted for joint fits.
The BAT emission is slightly more extended to the W-SW, as would be
expected from the higher temperature gas in that direction.
Note, however, that the outermost contour is consistent with noise
and should be ignored.
\label{fig:comabat:xmmbat}}
\end{figure}

The observations and processed \xmms data used herein are identical to
that presented in \citet{WSF+09}, where a more detailed description
can be found.
The \xmms EPIC-pn mosaic of Coma consists of 14 individual pointings,
the first set (11 pointings) of which were discussed in \citet{Bri+01}.
The full 14 pointings considered here were first presented in 
\citet{SFM+04}.
For joint fitting with \swifts BAT spectra, we extract events from a 
$65\farcm5\times65\farcm5$ box centered on the radio halo, which
was originally chosen to match the \suzakus Hard X-ray Detector field
of view; the region is shown as the outermost contour in Figure 3 of
\citet{WSF+09}, 
and also as the box in Figure~\ref{fig:comabat:xmmbat} in the present paper.
The only modification of our analysis procedure compared
to that in \citet{WSF+09} is that
the \xmms spectra were not weighted by the spatial response of the 
\suzakus HXD,
since they are not being fit simultaneously with that instrument.
No similar weighting is needed to comparison to the \swifts BAT data 
since the BAT survey covers the entire sky and the vignetting of
individual pointings is corrected for during processing.

% Figure 1

We also use the temperature map derived from the \xmms mosaic and 
described in \citet{WSF+09} to model the spatial distribution of 
hard X-ray emission, which is detailed in
Section~\ref{sec:comabat:spatial:models:thermal}.
Spectra extracted within a $16\times16$ contiguous grid of regions of size
$4\farcm3\times4\farcm3$ are fit to single temperature {\tt APEC} models
over the energy range $0.5 < E < 14$ keV.
While energies of $9.5 < E < 14$ keV are generally included during spectral
fits, these data are primarily used to determine the background and are
only included as a check of the background subtraction.
Because of the low count rate at these energies, their inclusion does not
significantly contribute to $\chi^2$ or the resulting best-fit parameters.

While data from the \suzakus observation of Coma (OBSID 801097010)
are not part of the current
analysis, we do make use of the 2--7.5 keV spectrum from the 0 chip of
the X-ray Imaging Spectrometer (XIS0) for calibration purposes, as
described in Section~\ref{sec:comabat:cn:xmm}.
We use the same spectrum that served to cross-calibrate the data in the
previous \xmm/\suzakus analysis \citep[region 10 from][]{WSF+09}.

\subsection{The \swifts BAT 58--Month Survey}
\label{sec:comabat:obs:bat}

The \swifts mission is primarily to detect and localize gamma-ray 
bursts,
which is accomplished with the very large FOV ($\sim 1/8^{th}$ of the sky), 
coded mask aperture Burst Alert Telescope (BAT).
As such, the nearly random pointing strategy culminates in an almost
uniform coverage of the entire sky with an $\sim 8$ Ms of exposure time,
made from many $\sim 5$ minute individual observations.
Images of the sky are reconstructed by cross-correlating the shadow
pattern of the randomly coded
mask in front of the detectors with a detector plane image via a fast
Fourier transform.
The detectors are sensitive to hard X-ray/soft gamma-ray photons from
14-195 keV in 80 native energy channels.
As part of the default survey processing, the channels are combined into
8 broader energy bands:
14-20 keV (E1), 20-24 keV (E2), 24-35 keV (E3), 35-50 keV (E4), 
50-75 keV (E5), 75-100 keV (E6), 100-150 keV (E7), and 150-195 keV (E8).
The final survey is built from the individual sky reconstructions,
which are summed and resampled onto predetermined image planes
of 6 facets, each in the Zenith Equal Area projection.
The detailed processing methodology and survey properties 
for the 58-month BAT all sky survey are nearly
identical to those described in \citet{Tue+10} for the 22-month survey.
The only major difference is that for the 58-month survey, the gain of each
detector pixel was {\it individually} calibrated with an onboard radioactive
source, which had not been done previously.
This better accounts for the sensitivity of low gain pixels,
effectively increasing the overall sensitivity.
Also, 
the sky images are more finely resampled so that the 
pixels near the center of the image projection scale $2\farcm8$
instead of $5\arcmin$\, as with previous versions of the survey.
The main advantage of this change is to improve the centroiding of
sources.

Because the systematic uncertainties in the survey-averaged spectrum of 
the Crab Nebula are smaller than the uncertainties in the BAT survey
response matrix, BAT survey fluxes are tied to the Crab fluxes in each
band \citep[see][Sec.~4.5]{Tue+10}.
One drawback to this approach is that a source flux is only guaranteed to be
correctly determined if its spectrum within the energy band is identical
the Crab's
(a power-law with a photon index of $\Gamma\sim 2.1$),
since the energy response within the band
may not be uniform.
Because we will fit the BAT spectrum jointly with the \xmms spectrum, 
it is important that the cross-calibration between the \swifts BAT 
and \xmms be accurate.
Since the flux calibration of the BAT survey is based on the Crab spectrum,
we have made sure that the \swifts BAT and \xmms agree on the flux 
and spectrum of the Crab.
The cross-calibration between \swifts BAT and \xmms is discussed in 
detail in Appendix~\ref{sec:comabat:cn}.

\subsubsection{The Coma Cluster}
\label{sec:comabat:obs:bat:coma}

The 6 facets of the BAT all sky survey
are oriented in Galactic coordinates with one facet centered
on each of the Galactic poles and the other 4 centered uniformly around
the Galactic plane.
The Zenith Equal Area projection conserves surface brightness but not
shapes, so objects far from the center of the projection can be
somewhat distorted.
However, the fortuitous location of the Coma cluster near the North
Galactic pole, and thus the center of its facet, means that any such
distortions are negligible.
Nevertheless, for all parts of the analysis image pixels are referred
to in terms of their Galactic coordinates so that any image projection
effects are completely eliminated.

\begin{figure}
\plotone{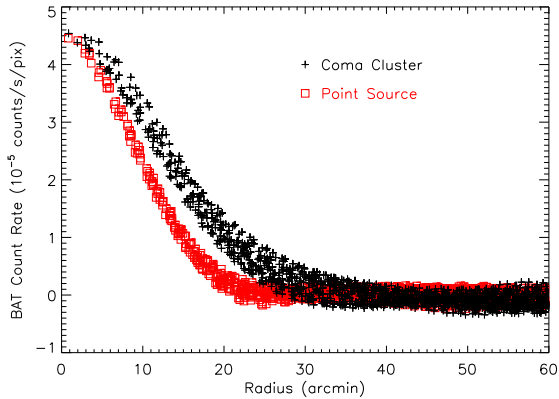}
\caption{The radial profile of the Coma cluster (crosses, black)
compared to a point source of comparable brightness (squares, red;
scaled slightly to match Coma's central flux).
Each point represents an individual pixel.
The BAT emission from Coma is clearly extended and not axially 
symmetric, as shown by the
the larger spread in pixel count rates in its profile compared to 
the point source.
\label{fig:comabat:comaext}}
\end{figure}

Hard X-ray emission is clearly detected in the first 4 energy bands, up to
50 keV.
In Figure~\ref{fig:comabat:xmmbat}, we present the wavelet-smoothed 2--7 keV
\xmms EPIC-pn image of the Coma cluster mosaic overlaid with contours
of the Crab-normalized BAT flux \citep[see][for a description]{Tue+10},
which shows the hard X-ray emission to be elongated in the same East-West
direction as the softer emission.
As first noted by \citet{Aje+09} in the BAT survey, the Coma
cluster is partially resolved by the BAT, which is explicitly shown
in Figure~\ref{fig:comabat:comaext}.
This figure compares the radial surface brightness profile of Coma 
with that of a nearby point source
with a comparable flux.
Coma is clearly extended.
The points plotted are individual pixels.
The greater width of the distribution for Coma indicates that its 
surface brightness is not circularly symmetric.
As shown in Figure~\ref{fig:comabat:xmmbat}, both the BAT and \xmms 
X-ray emission is elongated in a
ENE-WSW direction.

% Figure 2

While coded masks instruments have some difficulty detecting extended 
emission,
this is only true for emission extended on very large scales, 
when the shadow pattern
of the mask on the detector plane becomes sufficiently blurred.
Conceptually, as long as at least 1 detector pixel is
always shadowed by the mask over the entire extent of emission,
no flux can be mistaken for the unmodulated background.
The random placement of mask elements results in occasional,
continuous clumps of over 5 elements across, which means that
over a couple degrees all of the source flux can be recovered
since one mask element subtends, on average, about 20\arcmin.
For the BAT, we performed simple simulations that suggest that 
the actual scale over which accurate extended fluxes can be recovered, 
{\it in principle},
approaches the size of the FOV
if the spatial distribution of the source is known {\it a priori}.
The 
practical
problem is that flux from each part of an extended source
adds systematic noise to every other part of the source, eventually
drowning the signal in noise.
However, for small extensions this additional uncertainty is not
overwhelming.
In Appendix~\ref{sec:comabat:diffuse}, we show that the \swifts BAT 
provides accurate fluxes for extended sources on the scale of interest 
for the Coma cluster; 
in particular, Fig.~\ref{fig:comabat:diffsrc_recover} demonstrates that
all of the flux of a source extending over several degrees is 
accurately recovered.

Because nearby pixels in the BAT survey images are correlated, 
it is more difficult to determine the flux uncertainties for extended 
sources in the BAT than is usually the case for X-ray images.
For point sources, the flux uncertainties can be determined from the 
fluctuations in the local background in
the BAT survey images.
For Coma, we calculate the RMS fluctuations in the background 
($\sigma_{\rm bgd}$) around Coma in
an annulus of radius
$15<r<100$ pixels ($42\arcmin<r<4\fdg67$), as is typically done for 
sources in the BAT survey.
These values, and the flux uncertainties for extended sources like Coma, 
are derived in
Appendix~\ref{sec:comabat:err}.

%-------------------------- SPATIAL FITS --------------------------------
\section{Characterization of Extended Emission in BAT Images}
\label{sec:comabat:spatial}

To extract fluxes for extended sources, we choose to test {\it a priori}
model distributions, as opposed to using a method like the ``CLEAN''
algorithm \citep{Hog74}, which reconstructs fluxes from an unknown
underlining distribution assuming the PSF shape only.
(See Appendix~\ref{sec:comabat:diffuse:tests} for details about this choice.)
We represent a diffuse source as a collection of point sources, each
of which is convolved by the PSF (Eqn.~\ref{eq:comabat:psf}) and summed
together.
Throughout this work, image data are fit to these spatial models 
using the MPFIT algorithm \citep{Mar09}, 
which performs a Levenberg-Marquardt least-squares minimization
to converge on best-fit parameter values.

\subsection{Model Spatial Distributions of Hard X-ray Emission}
\label{sec:comabat:spatial:models}

Our goal is to detect IC emission from the same electrons producing
the radio halo; however, the electron spatial distribution need not
follow the radio halo if $B$ varies spatially within the ICM
\citep[e.g.,][]{DBD08}.
Indeed, there is evidence that the magnetic field in Coma declines
with radius \citep{BFM+10}, which would allow for a more extended 
relativistic electron population visible through IC interactions
with CMB photons.
To accurately search for this potential signal, we need to both choose
model distributions for this emission and to fully model the
thermal emission also present in the BAT energy bands.

\subsubsection{Thermal Models}
\label{sec:comabat:spatial:models:thermal}

Following the success of the \xmm-derived temperature
map for explaining the thermal origin of the
\suzakus HXD-PIN spectrum \citep{WSF+09}, we use the same map to
predict the spatial distribution of thermal emission at hard energies.
For each region of the map, the flux of the best-fit {\tt APEC} model is
calculated in each BAT energy band and treated as a point source at that
location.
Note that because the {\tt APEC} and {\tt MeKaL} 
look-up tables
are not 
defined above 50 keV
in {\tt XSpec}, we use {\tt MeKa} to derive the temperature map 
fluxes in the 4 highest
energy bands.
Then, for each band the 232 temperature map region ``points'' are taken
together to serve as the diffuse model.
To compare this or any of our diffuse models to the BAT image data,
each point is assigned the PSF shape with its peak value equal to the 
point flux, and the overlapping PSFs are summed together and sampled
at the location of the image pixels.

Thermal emission is detected in the first 4 BAT energy bands E1--E4.
Since this emission is an extension of the X-ray emission which dominates the 
\xmms image and its distribution is known, 
we fix the location of the thermal model to the best-fit
position of the model for the E1 band data, where the signal-to-noise
ratio is the highest.

Note that for the thermal model ``fits'' to the \swifts BAT 
spatial distribution in various bands,
only the normalization of the model in each band is allowed to vary.
The spatial distribution within each of the bands is completely 
determined from the \xmms data.

The \xmms data will also contain any nonthermal emission within 
the \xmms band.
Is it reasonable to use the \xmms temperature map and image to 
determine the spatial
distribution of the hard X-ray thermal emission?
The spectra in the temperature map were fit over the energy range 
0.5 keV$<E<14$ keV.
While the upper limit of this band is fairly hard, the low 
energy limit guarantees that the
spectra are dominated by softer photons.
For any sensible nonthermal spectrum, the \xmms spectra are 
dominated by thermal emission.
In fact, if there is cool, dense gas along a given line of sight, 
the \xmm-based model may actually underestimate
the thermal hard X-ray emission.
In any case, if there is strong nonthermal emission in the 
\xmms spectrum, it will dominate the
harder BAT energy bands, and will be uncovered in the joint 
spectral fits to the \xmms and BAT spectra
(Section~\ref{sec:comabat:specfits} below).

\subsubsection{Nonthermal Spatial Models}
\label{sec:comabat:spatial:models:nonthermal}

The \suzakus HXD-PIN upper limit, which is 2.5 times below the 
\rxtes and \saxs
detections, only excludes those measurements if the IC emission originates
from a relatively compact region ($R \la 20\arcmin$).
More extended emission of roughly uniform surface brightness, however,
would be consistent with both the detections and upper limit.
In most physical models for the IC, it is likely that the surface 
brightness would decline with radius;
one exception is the KW model discussed below.
However, there is no single well-established model for this decline.
Since our object is to test the possibility that the difference between our
\suzakus HXD-PIN upper limit and the \rxtes and \saxs detections is 
due to the extent of the IC emission,
we consider the extreme case of a uniform surface brightness disk.
Thus, 
we assume any nonthermal emission to take the form of 
a uniform brightness, circular disk with a radius $R = $ 25\arcmin\ (R25), 
30\arcmin\ (R30), 35\arcmin\ (R35), 40\arcmin\ (R40), 45\arcmin\ (R45), or
60\arcmin\ (R60).

Recently, \citet{KW10} proposed another model for the IC emission 
of the Coma cluster which
is consistent with both the \suzakus HXD-PIN upper limit and the 
\rxtes and \saxs detections.
In this model, the IC hard X-ray emission comes from a separate 
population of electrons from
those in the radio halo.
This new population of electrons are accelerated at the virial 
accretion shock of the cluster at
a very large radius.
These virial shock accelerated electrons lose energy quickly, and 
form a shell of hard X-ray emission,
which projects on the sky as a ring with nearly uniform surface 
brightness emission in its interior.
We will refer to this as the KW model.
While most of the flux resides in a ring at the cluster virial radius,
the amount of flux detected by an instrument pointed at the cluster center
will depend sensitively on its FOV.
We take all the model parameter values for Coma directly from \citet{KW10}
when comparing their expected flux to the constraints imposed by the
BAT data,
though the only relevant parameter for the spatial distribution is the
accretion shock radius $\theta_{200} = 82\farcm1$.
The radial distribution of flux is simply geometrical in form, assuming
an infinitely thin shell at this radius; the expression is given in
\citet[][Eqn.~9]{KW10}.

In reality, it is unlikely the spatial distribution of emission would
be as regular and axisymmetric as portrayed by these models.
However, for the spatial extent we consider relative to the resolution 
of the BAT, deviations from the idealized models will not particularly
impact our results.

We do not assume that the center of the nonthermal emission distribution 
from Coma is the
same as the center of the thermal emission.
Instead, for each of the nonthermal models, the model center is placed at
81 different positions on a $9\times9$ grid with $2\farcm5$ spacings
around the centroid of the large-scale thermal emission.
The center of the nonthermal emission is taken as one of the parameters 
to be varied
in the fits of the spatial and spectral distributions below.

\subsection{Spatial Fits to the BAT data}
\label{sec:comabat:spatial:fits}

The spatial models for the thermal emission alone, or for the thermal 
emission plus nonthermal emission, were fit to the pixel values in the 
BAT images in each of the 8 BAT bands.
As noted above (Section~\ref{sec:comabat:spatial:models:thermal}), the 
center of the thermal model was determined by fitting the center in the 
E1 band.
For the thermal model, only the overall normalization was allowed to vary.
For the nonthermal models, the center was varied for a grid of positions
(but fixed in each individual fit, see 
Section~\ref{sec:comabat:spatial:models:nonthermal}).
The model normalization (i.e., flux) in each of the 8 BAT bands was fit 
independently for the thermal and nonthermal models.
That is, the spatial fits made no assumptions about the spectrum of 
either type of emission.

\begin{figure*}
%\vskip-0.5truein
\plotone{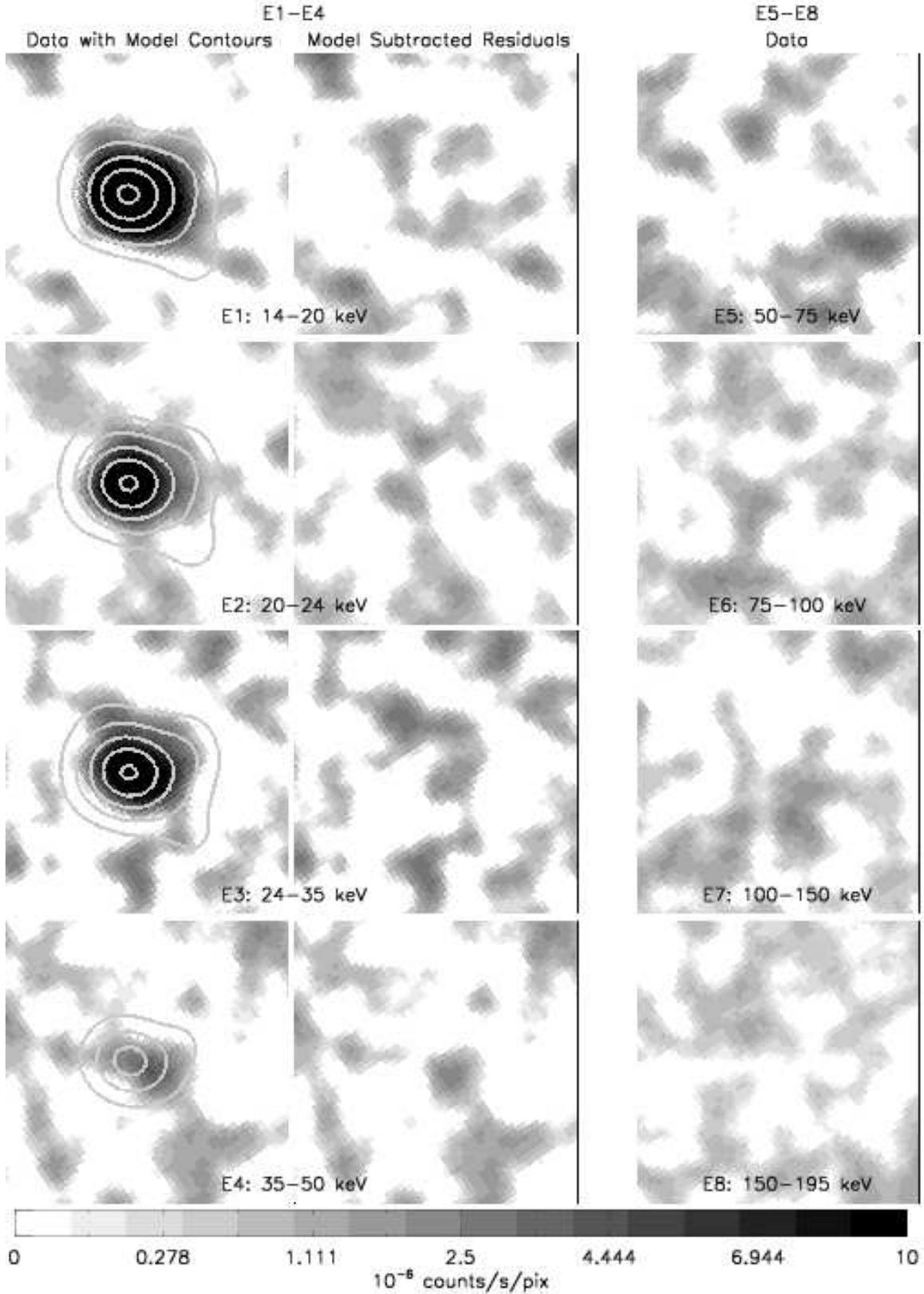}
%{0in}{0.}{340}{480}{60}{0}
\caption{Images from the 8 energy bands of the \swifts BAT survey (first and
third columns).
The greyscale follows a square root scaling from 0 counts/s
(white) to 
$>10^{-6}$ counts s$^{-1}$ pix$^{-1}$ (black).
The contours in the images in the first column show the best-fit 
thermal model for
each band, 
following a square root spacing.
For E1, the contours range from 0 to $4.2\times10^{-5}$ counts 
s$^{-1}$ pix$^{-1}$ and for E2 and E3, the contours range from 0 to 
$1.7\times10^{-5}$ counts s$^{-1}$ pix$^{-1}$, spanning 5 contours
in each case.
For E4, the 3 contours represent the model surface brightness at
0, $2.5\times10^{-7}$, and $10^{-6}$ counts s$^{-1}$ pix$^{-1}$.
The E1 contours occur at the same levels as shown in 
Fig.~\ref{fig:comabat:xmmbat}.
The middle column shows the thermal model-subtracted residual 
images for E1--E4,
with the same for greyscale as the data on the left.
The residuals show that the thermal spatial models are generally 
well-mapped to the actual data.
Note that the local background is also fit for and subtracted from the data
in the residual images, so the outer fluctuations are not identical 
to those in the left column.
Emission is clearly not detected in the 4 highest energy bands (E5-E8),
which are shown in the third column.
\label{fig:comabat:batimgs}}
\end{figure*}

\begin{figure*}
\plotone{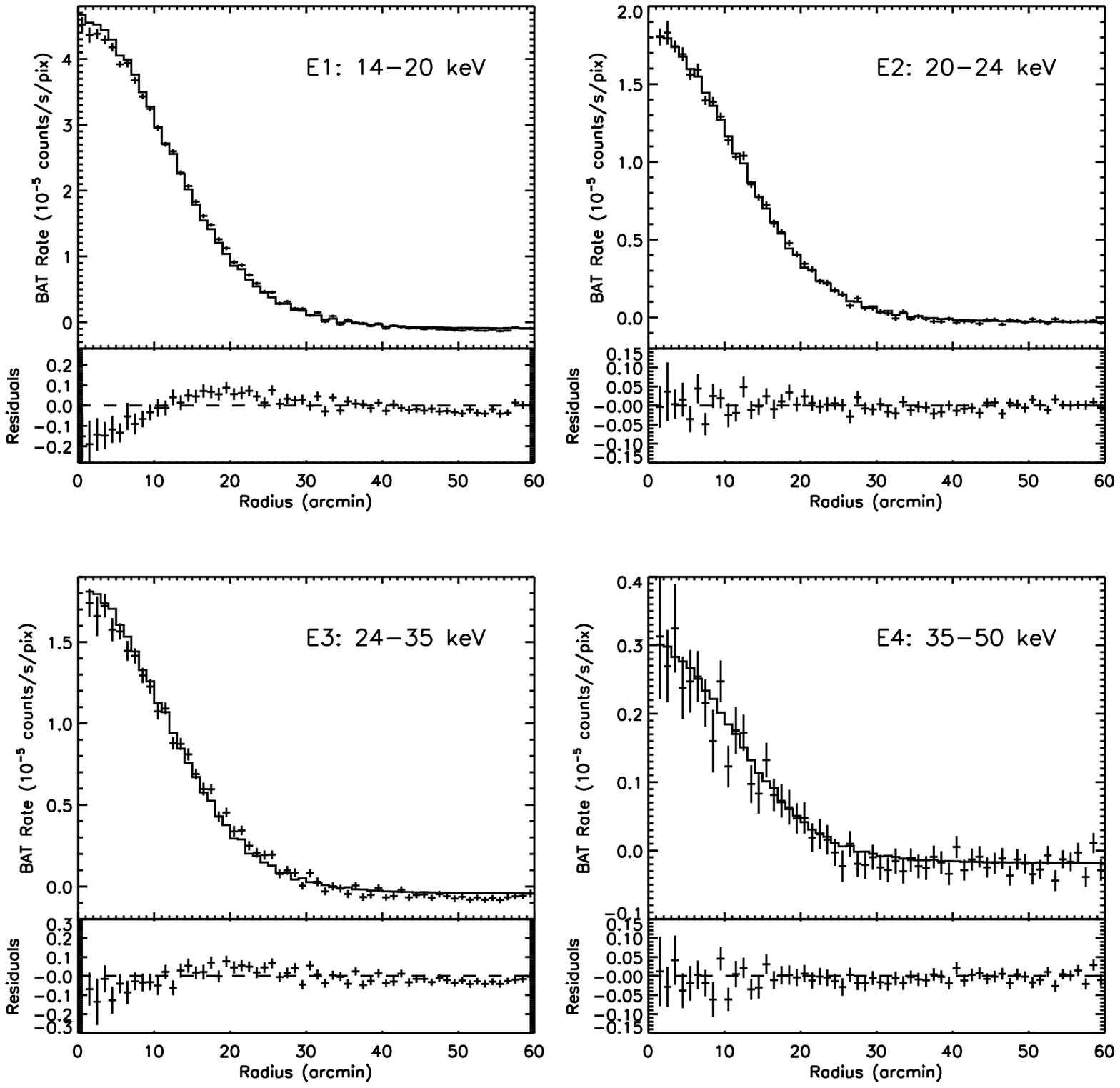}
\caption{Radial profiles of the data and
thermal model fits shown in the
first column images of Fig.~\ref{fig:comabat:batimgs}.
BAT pixels are averaged in annuli of 1\arcmin\ width (crosses), as are
the model values for each pixel position (histogram); the residuals
are plotted below each fit, on the same scale as the fit.
The structure in the E1 and E3 residuals could be due to a slightly
larger PSF FWHM and/or a true spatial distribution of emission that
differs slightly from our models; in either case, the effect on the
extracted flux would be less than its $1\sigma$ error.
\label{fig:comabat:proffits}}
\end{figure*}

In Figure~\ref{fig:comabat:batimgs}, we present the \swifts BAT images
of Coma in all 8 energy bands (first and third columns) along
with the thermal model-subtracted residuals for bands E1--E4 (center column).
The spatial distribution of the thermal models is represented with 
the contours
in the first column.
Note that only positively-valued pixels appear in the grayscale,
which has a square-root scaling, and that for each band pure black
corresponds to a slightly different value.
Each panel covers an identical $2\fdg7  \times 1\fdg5$ region of the sky.
Residuals from the fits are consistent with background fluctuations,
as can also be seen in the radial profiles shown in
Figure~\ref{fig:comabat:proffits}.

% Figure 3

% Figure 4

As noted above (Section~\ref{sec:comabat:spatial:models:thermal}),
these thermal models are solely based on the \xmms temperature map.
That is, the spatial distribution is completely determined from the 
\xmms data,
fit over the energy range 0.5 keV$<E<14$ keV,
and only the total normalization, independently measured in each
band, is allowed to vary.
Ideally, we would use a harder bandpass to derive the temperature
map, especially since it is known that such broad band single
temperature fits, when performed on a truly multi-temperature
spectrum, produces biased results \citep{NDG10}.
In practice, however, the lower count rate at energies above 2 keV
means that temperatures cannot be determined in regions with lower
temperature gas or lower flux.
We also extract such a temperature map, using the 2--12 keV band, and
find systematically higher temperatures, $\la10$ \% larger than the
temperatures in the map we use here.
Taking this map to represent the input spatial distribution,
we obtain statistically worse fits to the data, especially in the E1 band.
Because cooler, less bright gas in Coma's outskirts is not
included in this temperature map, though it is present in the BAT data,
the emission at larger radii cannot be properly fit.
In any case, the fluxes we derive in each BAT band do not significantly
depend on the absolute temperatures in the map, since only
relative temperature differences across the cluster matter, as the total
normalization is independently fit for in each band.
Because {\it all} the temperatures in the map are increased when a
higher energy bandpass is used, the flux in each BAT band remains
essentially the same regardless of which temperature map is adopted.
We therefore use the map determined from the 0.5 keV$<E<14$ keV band
in order to include the cooler gas at larger radius.
Regardless, this thermal model provides an adequate description of both
the spatial and spectral (discussed in 
Section~\ref{sec:comabat:specfits} below)
properties of the detected emission in the BAT data.
This success justifies our approach and confirms that extended emission
is detected with the same efficiency as that from the cluster center.

\begin{figure}
\plotone{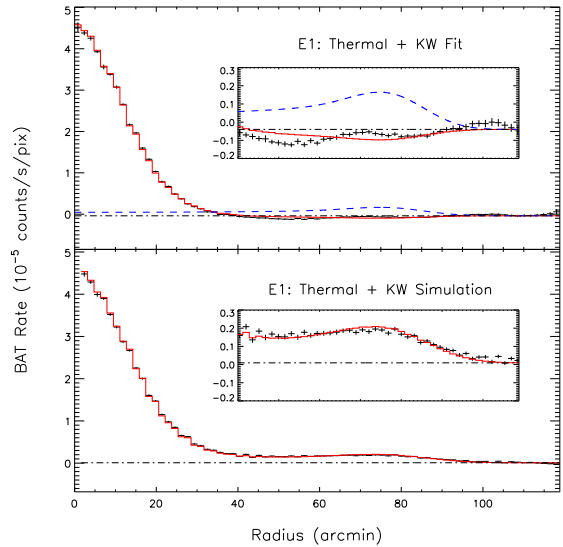}
\caption{The top panel shows the BAT E1 profile and spatial fit for a 
model with both
thermal emission and nonthermal emission following the \kws model 
(histogram, red).
The best-fitted value of the normalization of the \kws component 
is actually negative.
The dash-dot line represents the background level and the dashed line
(blue) shows the predicted spatial distribution of flux for the \kws model
\citep{KW10} from 14--20 keV.
The inset expands the scale of the y-axis above it to highlight the
difference between the data and the expected flux.
In the bottom panel, we perform the same fit to simulated BAT data 
based on the
thermal plus \kws model,
including shot and systematic noise comparable to that present in the
actual data.
This shows that the BAT would have easily detected a nonthermal component with
the spatial distribution given by the \kws model and the predicted flux.
\label{fig:comabat:thkwprof}}
\end{figure}

The good fit of the \xmm-based thermal model for the emission in
BAT bands E1--E4 and the
lack of obvious excess emission in the harder E5--E8 bands 
suggests that nonthermal emission is
not very strong or extended.
The nonthermal model with the most extended emission is the KW model 
\citep{KW10},
in which the IC hard X-ray emission comes from a thin shell at a 
very large radius.
Following the methodology in Section~\ref{sec:comabat:diffuse:tests},
we simulate the expected combined thermal and nonthermal flux
distribution for this model, and compare to the actual data.
The results for the E1 BAT band are shown in the upper panel of
Figure~\ref{fig:comabat:thkwprof}.
Clearly, we do not detect the nonthermal emission expected by
the \kws model.
The lower panel shows the thermal plus KW model compared to a 
simulation of the BAT data
assuming the distribution actually followed this model.
It is clear that the statistics in the BAT data would allow us to 
detect the nonthermal emission from the
KW model, were it present.
A more quantitative limit is derived in Section~\ref{sec:comabat:uls}.

% Figure 5

%-------------------------- SPECTRAL FITS ----------------------------
\section{Spectral Fits}
\label{sec:comabat:specfits}

To search for a nonthermal component in the X-ray spectrum of the
Coma cluster, spectral models are fit to the data.
The fits were done for the \swifts BAT spectrum alone, or simultaneously with
the \xmms spectrum of the cluster.
The BAT spectra were binned into 8 spectral channels, given by the 8 
standard BAT bands E1-E8.
To determine the spectrum in each channel, the total measured raw 
photon fluxes in each
band were converted into ``true'' photon fluxes using the 
calibration determined from the BAT Crab spectrum;
we also create a generic redistribution matrix to better represent
models with spectral shapes that differ from the Crab
(see Section~\ref{sec:comabat:cn:xmm}).
The uncertainties in each channel were determined from 
the flux uncertainty for a point source $\sigma_{\rm bgd}$
(see Section~\ref{sec:comabat:err:pterr}), and then corrected for 
the effects of source
extent as described in Section~\ref{sec:comabat:err:ext}, 
Equation~(\ref{eq:comabat:differr}),
and Table~\ref{tab:comabat:err}.
The final uncertainty is given by $\sigma_{\rm diffuse}$ in 
Equation~(\ref{eq:comabat:differr}).
Additionally, a problem with the implementation of the {\tt APEC},
which we use as our thermal description for spectral fitting, and 
{\tt MeKaL} models in {\tt XSpec} is that the 
look-up tables
are undefined above 50 keV.
Therefore, for the 4 energy channels above 50 keV, we substitute 
{\tt MeKa} for {\tt APEC}.
This should have no significant effect given the small thermal flux 
at these energies relative to the errors.

For the BAT-only spectra, in which only the nonthermal
component of the spatial fit is used to build the spectra,
a single power law model is sufficient to measure the nonthermal flux.
We also do joint fits of the \swifts BAT (containing both the thermal
and nonthermal spatial components) and \xmms spectra.
In these fits, the excellent statistics at low energies in the 
\xmms spectra very
strongly constrains the thermal emission.
However, the \xmms mosaic covers a smaller area compared to either
the \swifts BAT or most of the nonthermal spatial models.
(The \xmms extraction region used for this spectral analysis is indicated
by the square in Figure~\ref{fig:comabat:xmmbat}.)
Thus, in these fits the
models applied to the \xmms spectra are reduced by the fraction 
of the emission in our \xmms spectral extraction region.

\subsection{Joint \xmms EPIC-pn -- \swifts BAT Thermal Emission Fit}
\label{sec:comabat:specfits:thermal}

\begin{figure}
\plotone{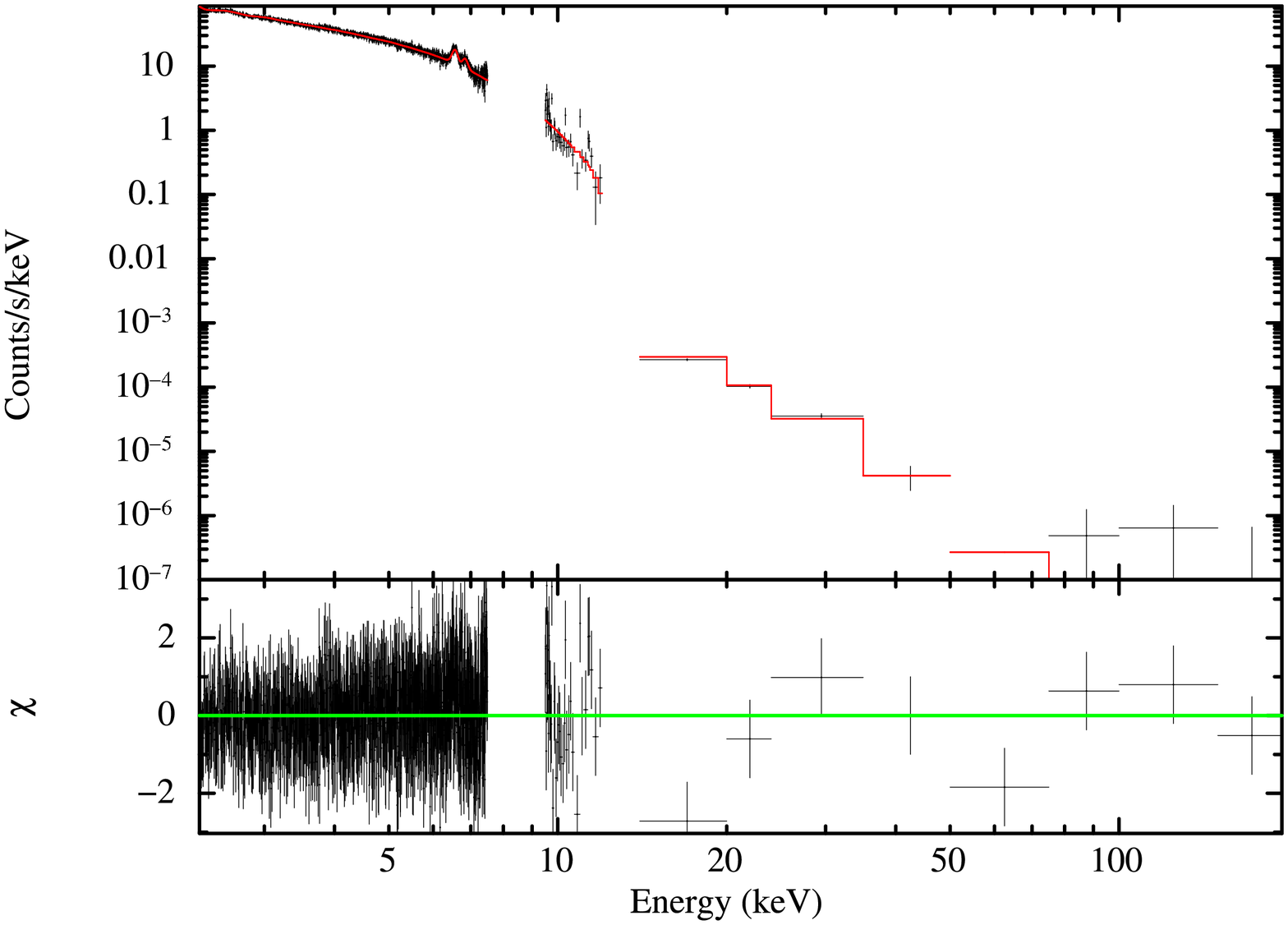}
%{0in}{-90.}{360}{500}{-20}{0}
\caption{Single temperature ({\tt APEC}, red histogram) 
fit to the \xmms EPIC-pn and \swifts BAT spectra.
The BAT spectra shown are reduced to the fraction which occurs in the 
\xmms spectral extraction region.
The BAT spectra were constructed assuming the spatial distribution
predicted by the \xmms temperature map.
The origin of the low E1 (14--20 keV) flux is discussed in the text.
A single temperature model ($kT = 8.24$ keV) is sufficient to
describe the 2--200 keV emission from the central square degree region
of the Coma cluster.
\label{fig:comabat:specth}}
\end{figure}

We first consider a purely thermal model for the X-ray emission in Coma, and
fit the \swifts BAT and \xmms spectra simultaneously.
The spatial distribution of the emission was assumed to follow the
thermal model (Section~\ref{sec:comabat:spatial:models:thermal})
as determined from the \xmms data.
For these and all following fits, we restrict ourselves to energies above
2 keV.
This choice ensures that the thermal component is not strongly biased
to lower temperatures by cooler gas in these global, multi-phase
spectra \citep[e.g.,][]{NDG10}.
The resulting single temperature fit is presented in
Figure~\ref{fig:comabat:specth}, and the parameters are given in 
the first row of
Table~\ref{tab:comabat:fits}.
The quality of the fit is quite good, indicating that a single component
description for the temperature structure is sufficient and that the
spectra have been reasonably well cross-calibrated.
A slight ascending trend in the E1-E3 residuals exists, however, which
is primarily due to a slightly lower than expected E1 flux.
While not particularly significant, it is worth mentioning several
potential causes for the trend.
The most straightforward explanation is that the calibration is slightly
wrong.
We presume the true Crab spectrum to be a simple power law across the
entire 2--200 keV interval, but if the spectrum actually steepens around
$E \sim 10$ keV as is likely the case \citep{Kir+05}, 
the higher energy bands will be assigned progressively 
higher flux
conversion factors (column 3 in Table~\ref{tab:comabat:err}); basically,
the flux in an energy band will be more and more overestimated for
bands at higher and higher energies.
Also, because emission is more significantly detected in the
lower energy bands,
a small change in the overall value of the \xmms and \swifts
cross-normalization factor -- such that the \swifts flux would be 
raised -- could reduce the spread in residuals.
From a more physical perspective, a single temperature model is not
entirely appropriate; in multi-temperature model fits, the trend in
residuals is not as strong.
In any case, an adjustment to the cross-calibration of 3--5\% is
sufficient to account for the trend, which is well within our
assumed 90\% confidence interval uncertainty of 10\%.

% Figure 6

%Table 1
\begin{deluxetable*}{llccccc}
\tablewidth{0pt}
\tablecaption{Joint Fits to  \xmms and \swifts Spectra 
\label{tab:comabat:fits}}
\tablehead{
Spatial &
Spectral &
$kT$ &
Norm.\tablenotemark{a} &
$\Gamma$ or $kT$ &
Norm.\tablenotemark{c} & $\chi^2$/dof \\
Model &
Model &
(keV) &
(cm$^{-5}$) &
\tablenotemark{b} &
&
}
\startdata
Thermal Region & Single T & $8.28\pm0.13$ & $0.373\pm0.002$ & - & - & 
1576.79/1544 \\
Thermal Region & 2T\tablenotemark{d}       & 7.8 & 0.25 & 9.4 & 0.12 & 
1575.77/1542 \\
Thermal Region & T$_{\rm map}$ & - & - & - & - & 1590.01/1545 \\ 
Thermal Region & T+IC\tablenotemark{e}     & $8.27\pm0.13$ & 
$0.373\pm0.003$ & 7.2 & 
$<$0.51 & 1576.79/1543 \\
Thermal Region & T+IC\tablenotemark{f}     & $8.27\pm0.13$ & 
$0.373\pm0.003$ & 2.0 & 
$<$0.00148 & 1577.21/1543 \\
Thermal \& KW & Single T & $8.30\pm0.13$ & $0.373\pm0.002$ & - & - & 
1570.32/1544 \\
Thermal \& KW & 2T\tablenotemark{e}       & 7.8 & 0.25 & 9.7 & 0.12 & 
1568.63/1542 \\
Thermal \& KW & T$_{\rm map}$ & - & - & - & - & 1584.59/1545 \\
Thermal \& KW & T+IC\tablenotemark{e}     & $8.30\pm0.14$ & 
$0.372\pm0.002$ & -1.5 & 
$<$$1.9\times10^{-9}$ & 1570.23/1543 \\
Thermal \& KW & T+IC\tablenotemark{f}     & $8.30\pm0.14$ & 
$0.373\pm0.003$ & 2.0 & 
$<$0.00082& 1570.62/1543 
\enddata
\tablenotetext{a}{Normalization of the {\tt APEC} thermal spectrum,
which is given by $\{ 10^{-14} / [ 4 \pi (1+z)^2 D_A^2 ] \} \, \int n_e n_H
\, dV$, where $z$ is the redshift, $D_A$ is the angular diameter distance,
$n_e$ is the electron density, $n_H$ is the ionized hydrogen density,
and $V$ is the volume of the cluster.}
\tablenotetext{b}{Value is $\Gamma$ for the T$+$IC model and $kT$ (in keV)
for the 2T model.}
\tablenotetext{c}{Value is the normalization of the power-law component
for the T$+$IC model, which is the photon flux at a photon energy of
1 keV in units of photons cm$^{-2}$ s$^{-1}$ keV$^{-1}$.
For the 2T model, the value is the normalization of the second {\tt APEC} thermal
model in units of cm$^{-5}$.}
\tablenotetext{d}{Parameters unconstrained.}
\tablenotetext{e}{Value of $\Gamma$ is fixed when deriving errors.}
\tablenotetext{f}{Value of $\Gamma$ fixed based on radio spectrum.}
\end{deluxetable*}

The best-fit temperature is $kT = 8.24 \pm 0.12 \, 
({\rm stat}) \pm 0.15 \, ({\rm sys})$ keV.
The systematic term in the error is based on varying the
cross-normalization factor by 10\%; the origin of this percentage
is discussed in Section~\ref{sec:comabat:uls} below.
This global temperature matches extremely well with previous
measurements.
For example, \citet{HBS+93} found $kT = 8.21$ keV with a spectrum
spanning $\sim2 < E < 11$ keV from the {\it Ginga} satellite, which
had a similar FOV (collimator with 1-2\arcdeg\ FWHM) to our aperture,
and \citet{Arn+01} found $kT = 8.25$ keV over a smaller FOV and
lower energy range ($0.3<E<10$ keV) with the \xmms EPIC-MOS instruments.
Including lower energy photons tends to lower single-temperature fits
to multi-component spectra \citep{CDV+08}, and having a smaller FOV
tends to emphasize the hotter central temperature of $kT\sim9$ keV in Coma,
which both explains why the {\it Ginga} and \xmms temperatures agree
and why the \xmm-\suzakus analysis of \citet{WSF+09} found a slightly
higher $kT=8.45$ keV.
While we find good agreement with other measurements, note that
our temperature, along with other temperatures derived with \xmm,
could be systematically cooler by a few tenths of a keV
than temperatures obtained with other observatories, given the steeper
\xmms Crab spectrum and the implications for its instrumental response
(see Section~\ref{sec:comabat:cn:xmm}).

We also tried a two-temperature thermal model for the \xmms and BAT data.
However, the two temperatures and normalizations
could not be individually constrained by the
data (Table~\ref{tab:comabat:fits}); the temperatures/normalizations 
listed in the table
result when the two-temperature model is fit for with initial 
temperatures of $kT_1=6$ keV and $kT_2=10$ keV.
The two-temperature fit is not a significant improvement on a 
single temperature fit.

While the average spectrum in the square degree region around Coma is
adequately described with one or two temperatures for the gas, in
fact the temperature distribution is quite non-isothermal.
We account for temperature variations in the spatial models used to
extract fluxes from the BAT images by extrapolating the \xmms
temperature map from \citet{WSF+09} to higher energies.
This map can also be converted into a spectral model (labeled 
``T$_{\rm map}$'' in Table~\ref{tab:comabat:fits}) and compared
to the joint spectrum.
By including the spatial information of the \xmms data in the
spectral model, we can better account for the thermal contribution
in the BAT energy bands.
The quality of this fit (allowing the normalization, but
not the shape, of the model to vary) is reported
in Table~\ref{tab:comabat:fits}.
In principle, the ``T$_{\rm map}$'' spectral model should perfectly
represent the total \xmms spectrum, but due to incomplete coverage
of the temperature map
with the total \xmms region and the imperfect determination of the 
individual temperatures, the $\chi^2$ value -- which is primarily
driven by the higher quality \xmms data -- is larger than for the other
fits, in which the model shape is free to vary and can account for these
minor differences.
Even so, the BAT data are slightly better described by this model than
by any of the other spectral models presented.
Although this result is perhaps expected, given that the ``T$_{\rm map}$''
spatial distribution is used to measure the BAT fluxes, it does
indicate that the method is self-consistent.

In all of the thermal models investigated, no evidence for
a strong high-temperature component is hinted at by the BAT data.

\subsection{Nonthermal Spectra}
\label{sec:comabat:specfits:nonthermal}

\subsubsection{Nonthermal Emission from the Cluster Center}
\label{sec:comabat:specfits:nonthermal:center}

To search for evidence of more centrally located nonthermal emission, 
an IC component was first fit to purely thermally-derived 
spectra -- i.e., spectra created from fits to the BAT data
using only the thermal spatial model -- which are reported in
rows 4 and 5 of Table~\ref{tab:comabat:fits}.
As in Section~\ref{sec:comabat:specfits:thermal}, all fits are to
the joint \xmm-\swifts spectrum extracted from the square region
in Figure~\ref{fig:comabat:xmmbat}.
Not surprisingly, the thermal model parameters are almost identical to
the fits without the IC component, and no significant IC emission is
present.
The good single temperature fit to the \swifts BAT and \xmms data 
already suggests that
the nonthermal contribution is not very significant.
If the photon spectral index $\Gamma$ of the nonthermal component 
is allowed to vary, it is unconstrained and the best-fitted value 
is unphysically steep, and in any case, only an upper limit can be 
placed on the nonthermal flux (Table~\ref{tab:comabat:fits}).
If we assume a photon index of 2.0 for the nonthermal component based 
on the radio data (Section~\ref{sec:comabat:uls} below),
the 90\% upper limit on the 20-80 keV flux is $1.24\times10^{-12}$
ergs/s/cm$^2$, which corresponds to $<0.8\%$ of the total flux in
the range $2<E<200$ keV.

\subsubsection{Extended Nonthermal Emission}
\label{sec:comabat:specfits:nonthermal:kw}

Our goal is to search for extended nonthermal emission, which is measured
with spatial model fits to the unlimited FOV BAT survey images
(Section~\ref{sec:comabat:spatial:fits}).
Because the thermal and nonthermal spatial model normalizations are
individually and simultaneously allowed to vary to best match the total 
flux in the BAT images, spectra can be created from the sum of
both components or separately, and also within any aperture.
For each of the 81 grid positions relative to the center of the cluster 
at which each nonthermal model was fitted for,
two spectra are produced.
One consists of the total, unvignetted flux of the nonthermal component
only, and the other includes the sum of both the
thermal and nonthermal emission inside the \xmms extraction region.
The second type of spectrum has the advantage
that it can be jointly fit with the EPIC-pn spectrum, which in the
case of a non-detection provides a tighter constraint on the flux
of nonthermal emission than the first type, since the nonthermal
component must be consistent with the higher quality,
lower energy \xmms data as well.
In none of these cases, for either type of spectrum, is a nonthermal
component detected with $\ga2\sigma$ confidence.
We therefore conclude that, while the \swifts BAT instrument is certainly
sensitive to extended emission, none of a nonthermal origin
is observed in the current version of the survey.
As an example, the fit parameters for various spectral models are shown
in Table~\ref{tab:comabat:fits} for the \kws spatial model
nominally positioned (i.e.\ centered on the large-scale \xmms emission).

\begin{figure}
\plotone{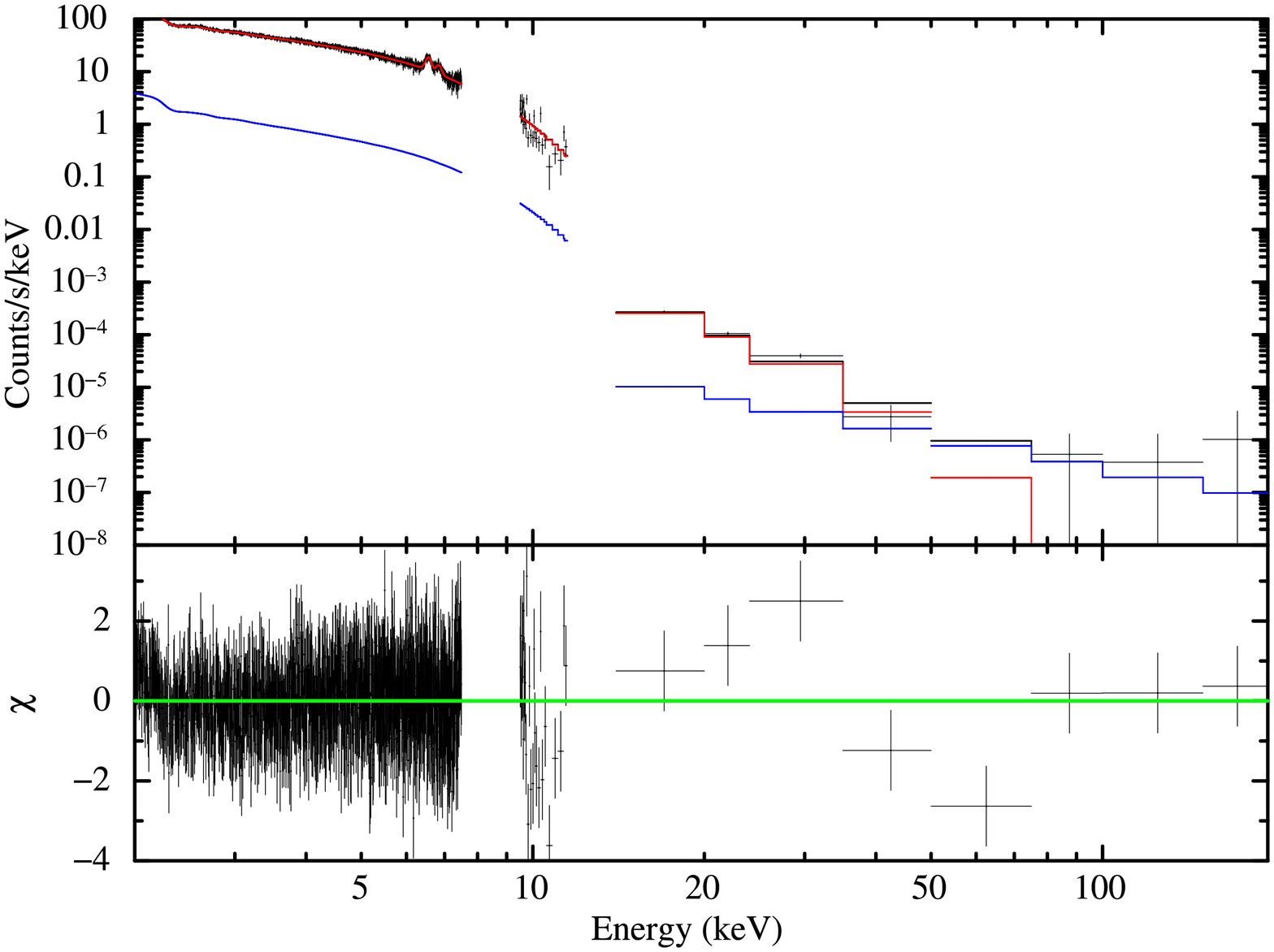}
%{0in}{-90.}{360}{500}{-20}{0}
\caption{
Thermal (red histogram) plus nonthermal (blue histogram) model 
fit to the \swifts and \xmms data for the 
$\Gamma=2$ power law nonthermal model corresponding to
the 90\% upper limit,
including the systematic uncertainties as described in the text.
This example, which is the model with most significant nonthermal flux,
is for a 25\arcmin\ radius, uniform surface brightness disk of 
nonthermal emission with a position offset
from the center of the large-scale thermal emission by
$-2\farcm5$ and 10\arcmin\, in $l$ and $b$, respectively.
As is true of all the upper limits from the joint
spectra, the nonthermal component does not exceed $\sim1\%$ of the
low energy part of the \xmms spectrum, nor does it compete with the
thermal emission until energies $E\ga50$ keV.
\label{fig:comabat:specul}}
\end{figure}

In Figure~\ref{fig:comabat:specul},
the nonthermal model with the most significant IC component is shown, 
assuming a fixed photon index $\Gamma=2$ for the spectral fits
of each nonthermal spatial distribution tried.
Note that the model in this figure represents the upper limit for a
nonthermal component, not its best-fit value, and that the
cross-normalization has been adjusted by 10\% in the direction that
favors a nonthermal signal.
The BAT spectra in Figures~\ref{fig:comabat:specth} and
\ref{fig:comabat:specul} are quite similar, indicating that even in
the most favorable case the data reject a significant IC contribution
to the spectrum of the Coma cluster, extended or otherwise.

% Figure 7

%-------------------------- UPPER LIMITS ------------------------------
\section{Upper Limits to Diffuse, Nonthermal Emission}
\label{sec:comabat:uls}

To ensure appropriate upper limits are derived, we must determine and
include any important systematic uncertainties in our results.
Typically, instrumental and/or cosmic backgrounds can be a serious
concern and must be carefully treated.
At lower energies where the thermal emission is bright, the background
is not comparable to cluster emission until an energy of $\sim7$ keV,
so even a background uncertainty of a few percent does not significantly
impact the \xmms EPIC-pn spectrum.
This background includes both the non-X-ray and cosmic backgrounds.
Given the size of the mosaic, the uncertainty in the overall EPIC-pn
background used here ($\sim2$-12 keV) is 2.4\% \citep{RP03}.
Increasing the \xmms background favors larger nonthermal fluxes since 
this reduces the temperature slightly, so the
\xmms background is raised by 2.4\% during upper limit derivations.
Point sources in the \xmms mosaic account for only $\sim1\%$
of the emission and have a spectrum that as a whole does not vary
significantly from a $\Gamma\sim2$ power law \citep{WSF+09}, so we
do not model their contribution to the \xmms spectrum.
While their flux may artificially enhance a nonthermal signal, 
ignoring them will only result in slightly more conservative upper limits.
For the BAT survey data, the background is automatically removed as
part of the image reconstruction procedure, and systematic variations are 
encoded as fluctuations in empty sky regions, which is already included
in the error budget.

The more significant systematic uncertainty is in the determination of
the cross-calibration between \swifts and \xmm.
Ideally, there should be no uncertainty since we based the BAT
calibration on the \xmms data and the \suzakus XIS0 Crab spectrum.
However, the slope of the calibration (i.e.\ the assumed photon index of the
Crab) is less certain.
The total error, statistical and systematic, of the photon index in
\xmms EPIC Crab fits is $\pm 0.05$ \citep{Kir+05}, so we adjust our
calibration to make the canonical Crab spectrum flatter by 0.05, which
acts to increase the BAT fluxes ($\sim10\%$ for E1, $\sim20\%$ for E8)
and flatten the BAT spectra, thus enhancing nonthermal fluxes.
This approach to the systematic uncertainty is also conservative, 
as it is known that
the Crab spectrum steepens above 10 keV \citep[e.g.,][]{Kir+05}.

Since we have no clear detection of nonthermal emission, we must
decide on its photon index from other arguments.
The natural choice is to use the spectral index of the radio halo,
or $\Gamma=$ 1.5--2.5 \citep{GFV+93}, 
though the lower range of the X-ray regime
explored here corresponds to lower energy electrons where the emission
may have a flatter spectrum.
Also, both previous detections using \rxtes and \saxs data found 
$\Gamma\sim2$, and the model of \citet{KW10} predicts this photon
index.
Therefore, we fix the nonthermal power law index to $\Gamma=2$, 
primarily because we are most interested in directly comparing our
upper limits with these previous detections and model predictions.
If the spectrum of nonthermal emission is in fact flatter, our upper
limits will be low by some amount since the BAT errors are large and
the \xmms data will have less leverage on the fits.
However, the high energy flux will not increase dramatically; as
illustrated in Table 3 of \citet{WSF+09}, the 20-80 keV flux rises 
by a factor of 2 from $\Gamma=2$ to
$\Gamma=1.5$, and trials
show the same behavior for the nonthermal component in this work.

\begin{figure*}
\plotone{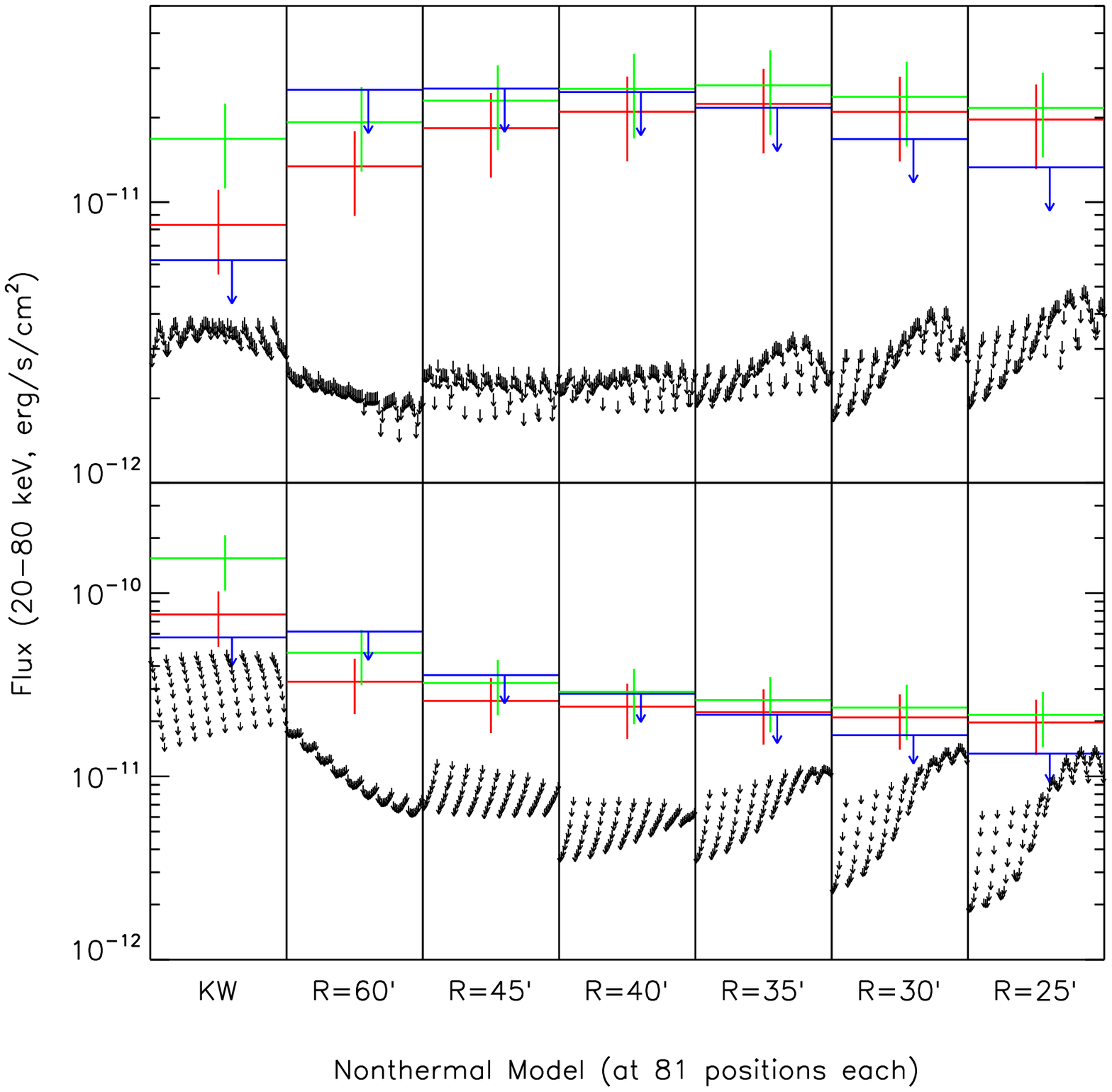}
\caption{Upper limits (small arrows) for each nonthermal spatial model
relative to the \rxtes \citep[][upper cross, green]{RG02}, 
\saxs \citep[][lower cross, red]{FOB+04}, and 
\suzakus \citep[long arrow, blue]{WSF+09} detections/upper limit.
The previous flux detections [$(1.5 \pm 0.5)\times10^{-11}$ ergs/cm$^2$/s]
and upper limit ($6\times10^{-12}$ ergs/cm$^2$/s) are corrected to
account for the fraction of emission missed due to vignetting by the
collimator response functions (see text).
In the top panel, upper limits are calculated from the simultaneous
joint fits to the \xmms and \swifts spectra, and all nonthermal fluxes
reported are from inside the \xmms extraction region (the square in
Fig.~\ref{fig:comabat:xmmbat}).
In the bottom panel, upper limits are derived from BAT spectra created 
from the nonthermal component of spatial fits {\it only}, and the fluxes
represent the total emission of the spatial model.
Based on the results presented in the top panel, we conclude that
extended IC emission cannot reconcile the discrepancy between the
\suzakus and \rxte/\saxs observations.
\label{fig:comabat:ul}}
\end{figure*}

For the above systematic uncertainties, we find the 90\% confidence
upper limits to nonthermal emission for both the nonthermal-only BAT
and for the joint EPIC-pn/BAT spectra.
The thermal component in the latter case is simultaneously fit
with the normalization of the nonthermal component.
We present each individual limit in Figure~\ref{fig:comabat:ul}
along with the vignetting corrected fluxes/upper limit from
\rxtes \citep[][upper cross, green]{RG02}, 
\saxs \citep[][lower cross, red]{FOB+04}, and 
\suzakus \citep[][upper limit, blue]{WSF+09}.
The collimator responses for these instruments are convolved with 
the model flux distributions to give these values or limits.
The \suzakus HXD-PIN instrument is a square collimator with spatial
sensitivity of the form given in Equation~3 of \citep{WSF+09}, and
the \rxtes PCA/HEXTE and \saxs PDS instruments are hexagonal collimators
with triangular approximation FWHM of 1\arcdeg\ and 1\fdg3, 
respectively.
We approximate the nearly axisymmetric response with a 4th-order
polynomial of the form:
\begin{equation} \label{eq:comabat:collhexresp}
\begin{array}{rcl}
R_{\rm hex}(\theta) & = & 1.00 - 
1.36 \left(\frac{\theta}{\theta_{\rm max}}\right) + 
0.46 \left(\frac{\theta}{\theta_{\rm max}}\right)^2 \\ & & - 
0.58 \left(\frac{\theta}{\theta_{\rm max}}\right)^3 + 
0.48 \left(\frac{\theta}{\theta_{\rm max}}\right)^4
\end{array}
\, ,
\end{equation}
where $R_{\rm hex}(\theta)$ is the fraction of emission visible to the
instrument at off-axis angle $\theta$ and $\theta_{\rm max}$ is where
emission is no longer detected.
For each spatial model, our upper limits are ordered in Galactic
coordinates from the lowest values of $l$ and $b$ in our grid,
incrementing $l$ for all positions with that latitude before
incrementing $b$, with $l$ reset to the minimum value.
It is this ordering that produces the pattern evident in the limits.
The limits for the joint spectral fits are given in the top panel, 
while the nonthermal-only spectral limits are provided in the bottom
panel.

%Figure 8

Surprisingly, the upper limits derived from extended spatial models
and the BAT data alone (bottom panel) are comparable in sensitivity 
to the previous detections/limit.
Larger models are generally less constrained, due to the greater
uncertainty in their estimated flux (Equation~\ref{eq:comabat:differr}),
though local fluctuations have a greater impact on smaller models,
increasing the spread with position.
Stronger constraints are obtained when lower energy emission is
simultaneously considered (top panel), and a similar range in upper
limits is found for each spatial model distribution.
This result is not surprising, as each model contributes roughly the
same amount of flux inside the \xmms extraction region, since most of
them are extended beyond this region.
Interestingly, the \kws model provides the limits most consistent with
the \saxs detection, which follows from the large PDS FOV -- it would
observe a higher proportion of the brighter ring emission -- and the
larger errors resulting from the \kws model's size.
However, in all realizations of the joint fit case, our 90\% limits
lie below the 90\% interval of the previous detections/limit.
Thus, all of the previous detections are excluded for any of the 
spatial models when
one fully accounts for the differences in spatial sensitivity 
between instruments.
Our upper limits, for the nominal case where the nonthermal distribution
is centered on the large-scale thermal emission, are compared to
these previous measurements in Table~\ref{tab:comabat:uls}.

%Table 2
\begin{deluxetable}{lcccc}
\tablewidth{0pt}
\tablecaption{Flux Upper Limits (20--80 keV) for Nominal NT 
Position\tablenotemark{a}
\label{tab:comabat:uls}}
\tablehead{
Model &
Joint BAT limit &
\sax &
\rxte &
\suzaku \\
& $10^{-12}$ cgs &
$10^{-12}$ cgs &
$10^{-12}$ cgs &
$10^{-12}$ cgs
}
\startdata
\kw & $<$ 3.86 & $16.8 \pm 5.6$ & $\phn8.3 \pm 2.8$ & $<$ \phn6.2 \\
R60 & $<$ 2.16 & $19.3 \pm 6.4$ & $13.4 \pm 4.5$ & $<$ 25.1 \\
R45 & $<$ 2.34 & $23.0 \pm 7.7$ & $18.4 \pm 6.1$ & $<$ 25.4 \\
R40 & $<$ 2.53 & $25.3 \pm 8.4$ & $21.0 \pm 7.0$ & $<$ 24.7 \\
R35 & $<$ 2.95 & $26.1 \pm 8.7$ & $22.4 \pm 7.5$ & $<$ 21.7 \\
R30 & $<$ 3.22 & $23.7 \pm 7.9$ & $21.0 \pm 7.0$ & $<$ 16.8 \\
R25 & $<$ 3.48 & $21.7 \pm 7.2$ & $19.7 \pm 6.6$ & $<$ 13.3 \\
\enddata
\tablenotetext{a}{Units for flux are $10^{-12}$ erg cm$^{-2}$ s$^{-1}$}
\end{deluxetable}

%-------------------------- DISCUSSION --------------------------------
\section{Implications and Discussion}
\label{sec:comabat:disc}

By taking advantage of the crude imaging capabilities of the \swifts
BAT instrument and the impressive sensitivity of the 58--month all
sky survey, we are able to constrain the amount of nonthermal,
hard X-ray emission -- extended or otherwise -- from the Coma
cluster.
We find no evidence for an extended, hard excess that could reconcile
recent detections from \rxtes \citep{RG02} and \saxs
\citep{FOB+04} with the upper limit from \suzakus \citep{WSF+09};
note, however, these detections would still be in conflict with
the upper limit of \citet{RM04}.
Generic, uniform surface brightness disks, along with a recently
proposed IC model \citep{KW10}, were fit to BAT
survey images, converted to spectra, and investigated for signs
of a nonthermal component.
For each spatial model, we compute upper limits on a grid of
positions and compare them to previous measurements, being careful
to convert detected fluxes into intrinsic source fluxes, given a
particular spatial distribution, by
accounting for the collimator vignetting functions.
These are direct comparisons, in the sense that the instrumental
response of all detectors involved have been fully considered,
and as such we, like \citet{RM04}, cannot confirm the claimed
detections of \citet{RG02} and \citet{FOB+04}.

Their observed hard excesses could have had other reasonable
sources, if not diffuse IC emission from the nonthermal phase of
the ICM.
A common difficulty is an accurate determination of both the cosmic
and non-X-ray background, the treatment of which is the primary
difference between \citet{FOB+04} and \citet{RM04}.
Another possibility is the variable nature of nearby point sources,
most notably the AGN W Comae, which was once quite bright but has
been fading for many years.
The concurrence of the \rxtes and \saxs observations could have
simply been unlucky and caught W Comae (or another source) in
a bright state.
A somewhat more subtle, and perhaps more likely, explanation concerns
the multi-temperature nature of Coma's ICM.
Small amounts of hot gas could
dominate the high energy emission, so the
extrapolation of an average temperature determined from lower
energy data may not be an adequate description of the thermal
contribution to the high energy flux.
The effect of the multi-temperature gas in Coma is evident in the 
SW extension at hard energies
observed by \integrals \citep{RGL+06, ENC+07} and confirmed here;
higher temperatures seen at this location in the temperature
map \citep{WSF+09} are sufficient to explain the change in morphology,
which points to the increased significance of this gas at higher energies.
Even so, single temperature spectral fits do not produce IC detections
in this study or in \citet{WSF+09}.
The explanation may simply rest in a slight mischaracterization
of the hard energy emission weighted temperature; in \citet{FOB+04},
the FOV of the lower energy HPGSPC instrument does not quite match the
higher energy PDS, and the temperature of $7.67\pm0.1$ keV found in 
\citet{RG02} is significantly below that allowed by the \xmms
data.

Given radio synchrotron emission and an upper limit on the
X-ray IC flux, a lower limit on the average ICM magnetic field
can be estimated, as described by equation~(13) in \citet{WSF+09} and the
accompanying text.
A diffuse radio flux of 640 mJy at 1.4 GHz is detected out to a radius of 
$\sim40\arcmin$ in \citet{DRL+97}.
For comparison, we will use the 
upper limit of $2.7\times10^{-12}$
erg s$^{-1}$ cm$^{-2}$
($20 < E < 80$ keV)
from the $R=40\arcmin$ disk model.
These values imply
$B>0.25$ $\mu$G,
an increase from
\citet{WSF+09} but still well below the equipartition value 
of $B_{\rm eq}=0.5$ $\mu$G for the Coma radio halo \citep{GFV+93}.
A slightly lower limit of $B>0.2$ $\mu$G results if a more conservative
IC upper limit of $4.2\times10^{-12}$ erg s$^{-1}$ cm$^{-2}$ is used,
which considers the limits from all spatial models tested.
Regardless, these limits on $B$ fall well below line of sight
estimates of several $\mu$G from Faraday 
rotation measure (RM) observations \citep{FDG+95}, though due to
geometric effects these measurements may not represent the average
cluster magnetic field \citep{Pet01}.

However, the global field may be recovered by combining many RM
measurements along different lines-of-sight through the ICM with 
numerical simulations \citep{Mur+04}.
\citet{BFM+10} have applied this method to the Coma cluster,
deriving a radial profile where the energy density of the magnetic
field falls roughly in proportion with the energy density of thermal
gas and with a central field strength of $B_0 \sim 4.7$ $\mu$G.
Combining this model of $B(r)$ with an approximate representation
of the radial density profile of synchrotron emission, implied by
a rough $\beta$-model fit to the point source subtracted image of
\citet{DRL+97} ($r_c=18\arcmin$, $\beta=1$, 
and $I_0=1.23$ mJy arcmin$^{-2}$), directly leads to a prediction of
the expected IC surface brightness as a function of radius.
Our illustrative -- due to the large uncertainties in all parameters
assumed in this exercise -- IC surface brightness distribution
is flat out to $\sim 30\arcmin$ with a 20--80 keV flux of
$\sim 8\times10^{-17}$ erg s$^{-1}$ cm$^{-2}$ arcmin$^{-2}$, 
at which point it nearly linearly drops
toward zero, though not reaching it, around a radius of 90\arcmin.
This surface brightness is about an order of magnitude below that
implied by our upper limits, providing a possible explanation for
why we are unable to detect an IC signature.
An order of magnitude lower hard X-ray flux for Coma is also predicted by 
\citet[][see their Fig.~5]{BL10},
who have developed the most comprehensive picture yet of radio halo
generation by MHD turbulence.
On the other hand, larger IC fluxes would be expected if the radio
synchrotron emission falls off more gradually than modeled here,
since a flatter radial profile would suggest a higher relativistic
electron density given the falling magnetic field with cluster radius.
More accurate maps of Coma's radio halo, preferably at 
lower frequencies where the radio electrons correspond more closely to
the IC-emitting electrons, will clarify this issue.

Ultimately, a true detection of IC emission from Coma will have to
wait for upcoming missions with focussing hard X-ray telescopes,
namely {\it NuSTAR}\footnote{http://www.nustar.caltech.edu/} and
{\it Astro-H} \footnote{http://astro-h.isas.jaxa.jp/}.
For {\it NuSTAR} to achieve a sensitivity comparable to our upper 
limits, a single pointed observation of at least 100 ks will be
required \citep{MHK+09}.
However, the much finer spatial resolution will remove the
uncertainty associated with bright background AGN and
allow multiple spatially-resolved joint fits.
Assuming the hottest gas, which produces the largest amount of thermal
emission at hard energies, is localized, then these regions can be 
identified and avoided in order to detect a lower surface 
brightness, but more uniform, IC component.
Similarly, if the IC emission is more localized, it will be easier to
identify with spatially-resolved joint fits between \xmms and
{\it NuSTAR} or {\it Astro-H} spectra, as has been done with
{\it Chandra} data alone \citep{MA09}.
The unambiguous detection of IC emission associated with radio halos
and relics is crucial to determining the energy content
in the relativistic phase of the ICM and how significant of an
influence this phase has on the dynamics and structure of the
thermal gas in clusters.

%-------------------------- ACKNOWLEDGEMENTS --------------------------
\acknowledgments
We thank W.\ Reich who kindly provided us with the \citet{DRL+97}
radio image, C.\ B.\ Markwardt who explained to us (and wrote) many of
the BAT software analysis routines used in this work, and the referee
whose comments helped clarify the discussion of several key points.
DRW was supported by a University of Virginia GSAS Dissertation Year 
Fellowship and a Virginia Space Grant Consortium Fellowship.
DRW and CLS were supported in part by NASA through Suzaku grants 
NNX08AZ99G, NNX09AH25G, and NNX09AH74G, and XMM-Newton grants 
NNX08AZ34G and NNX08AW83G.
Basic research in radio astronomy at the NRL is supported by 
6.1 Base funding.

%---------------------------- APPENDICES ------------------------------

\appendix

%---------------------------- CALIBRATION -----------------------------
\section{\xmms EPIC-pn--\swifts BAT Cross-Calibration}
\label{sec:comabat:cn}

Because the systematic uncertainties in the survey-averaged spectrum of 
the Crab Nebula are smaller than the uncertainties in the BAT survey
response matrix,
BAT survey fluxes are tied to the Crab fluxes in each
band since the systematic uncertainties in the survey-averaged spectrum of 
the Crab Nebula are smaller than the uncertainties in the BAT survey
response matrix
\citep[see][Sec.~4.5]{Tue+10}.
This method also requires that the intrinsic Crab spectrum be defined since
its exact spectrum remains somewhat uncertain 
(see Sec.~\ref{sec:comabat:cn:hard} below), particularly at higher 
X-ray energies.
In practice, though, we are less concerned with an accurate absolute
calibration
for the BAT than we are with, in this case, an accurate calibration
{\it relative to} our \xmms EPIC-pn spectrum.
Therefore, instead of prescribing a canonical Crab spectrum as close to
the true spectrum as it has been measured thus far, we need to set it
to the Crab spectrum as measured by the \xmms EPIC-pn instrument
{\it over the energy range we consider}.
Otherwise, systematic calibration errors between the instruments could
significantly affect our result, since our goal is to detect excess
radiation at hard energies due to nonthermal emission.
Errors leading to steeper (flatter) \xmms spectra and flatter (steeper) 
\swifts spectra, for example, will reduce (increase) the thermal
contribution at higher energies and similarly enhance (suppress) a
nonthermal signal.
In other words, any systematic miss-calibrations are mimicked in the
BAT calibration so that thermal and nonthermal models can be simply
applied during joint fits of the data.

\subsection{The Spectrum of the Crab According to \xmm} 
\label{sec:comabat:cn:xmm}

Because of its high X-ray flux, simple spectrum, and lack of 
significant variability, the pulsar wind nebula of the Crab 
supernova remnant has been proposed as an X-ray standard flux
calibrator \citep{Kir+05}.
Observations over a large range of energies and with many diverse
instruments reveal an intrinsic spectrum nearly consistent with a 
single power law; however, the photon index and normalization
determined by each detector exhibit small but not insignificant
differences \citep{Kir+05, WGJ+10}.
\xmms EPIC-pn measurements, which require that observations are made
in burst mode due to \xmm's large collecting area,
are best fit with a steeper than average photon index 
\citep[$\Gamma=2.13$ versus $\langle \Gamma \rangle =2.08$ for 
simultaneous fits to many
instrument observations,][]{Kir+05}
that is driven 
by the shape of the spectrum from 0.5-2 keV.
However, the residuals to this fit in their Figure~7 suggest that
for energies above 2 keV, the \xmms photon index is more in line
with the average, and since we only consider energies above 2.3 keV,
we need to determine what the Crab spectrum is measured to be in this
range.

Instead of fitting the Crab spectrum directly, we choose to compare
the \xmms data to \suzakus XIS0 data, which has been well calibrated
and consistently fit over its energy range using observations of the Crab.
Spectra from each instrument are extracted from identical spatial regions 
at the center of the Coma cluster
\citep[specifically, Region 10 from][]{WSF+09},
where both the gas temperature and surface brightness are roughly
constant.
Fitting each spectrum from 2.3-12 keV 
with a single temperature {\tt APEC} model, we find the
\xmm-derived temperature to be slightly, though not insignificantly,
lower than the \suzakus temperature: 8.32 keV versus 8.90 keV.
The lower EPIC-pn temperature is consistent with an increasingly larger
effective area at higher energies relative to \suzaku's high energy
effective area; positing that \xmm's calibration is correct,
the relative \suzakus effective area at larger energies should be 
increased as a function of energy, which would lower the flux and
therefore the temperature.
We model this effect as multiplicative power law component to the 
\suzakus {\tt APEC} fit, which simultaneously accounts for both the gradient
and the overall cross-normalization between the two instruments.
Fixing the {\tt APEC} model parameters to those found with the fit to the
\xmms spectrum, we find this modification to the \suzakus calibration:
$f (A_{eff}) = 0.923 (E/1 \, {\rm keV})^{0.045}$.
In other words, \suzakus spectra are 
flatter than \xmms spectra and have similar hard band fluxes; 
however, note that the \xmms effective area
had previously been reduced by 15\%, as per the analysis in \citet{WSF+09},
in order to match the 2-10 keV EPIC-pn and XIS0 fluxes.
Dividing the XIS0 best-fit Crab spectrum of 
Ishida et al. [$F_{\rm Crab,XIS0}(E) = 9.51 (E/1 \, {\rm keV})^{-2.05}$, 
given in
{\it Suzaku} Memo 2007-11\footnote{http://www.astro.isas.ac.jp/suzaku/doc/suzakumemo/suzakumemo-2008-03.pdf}]
by $f (A_{eff})$ finally yields the correct parameters for the \xmms
fit to the Crab spectrum in the energy range of interest ($E > 2$ keV):
\begin{equation} \label{eq:comabat:crab}
F_{\rm Crab}(E) = 10.30 \left(\frac{E}{1 \, {\rm keV}}\right)^{-2.095} 
{\rm photons/cm^2/s} 
\, .
\end{equation}
We take this equation as our canonical Crab spectrum.
Then, the true flux of a source in each of the 8 bands is given by 
the BAT source count rate in that band divided by the observed Crab BAT 
count rate in that band, and multiplied by the spectrum in
Equation~(\ref{eq:comabat:crab}) integrated over the band
\citep[see][eqns.~2--4]{Tue+10}.
This conversion factor is reported in Table~\ref{tab:comabat:err} as 
the ``Flux Calib.''.
Note that this photon index differs significantly from that assumed
in \citet{Tue+10}, $\Gamma=2.15$.
Our goal here is to choose a photon index that agrees with the \xmms
data and best matches the Crab spectrum at the lowest BAT energies,
where it is flatter than at harder energies (see the discussion in
Section~\ref{sec:comabat:cn:hard}).

%Table 3
\begin{deluxetable}{lccccccccccccc}
%\rotate
\tablewidth{0pt}
\tablecaption{\swifts BAT Error Factors  \label{tab:comabat:err}}
\tablehead{
&
Energies &
Flux Calib.\tablenotemark{a}&
$\sigma_{\rm bgd}$\tablenotemark{b} &
PtSrc &
\multicolumn{2}{c}{Thermal} &
R25 &
R30 &
R35 &
R40 &
R45 &
R60 &
\kw \\
Band &
(keV) &
(cm$^{-2}$) &
&
$f_m$ & $N_{\rm PSF}$ & $f_m$ & \multicolumn{7}{c}{$f_m$} 
}
\startdata
E1 & \phn14--20\phn & 16.55 & 2.37 &  1.44 &   2.09 &  1.80 &  1.77 &  
1.60 &  1.73 &  \phn1.98 &  \phn2.08 &  \phn2.06 &  \phn2.22 \\
E2 & \phn20--24\phn & 11.28 & 1.05 &  1.32 &   2.02 &  1.75 &  1.39 &  
1.53 &  1.56 &  \phn1.22 &  \phn1.60 &  \phn1.57 &  \phn1.94 \\
E3 & \phn24--35\phn & 10.89 & 1.63 &  1.35 &   1.97 &  1.59 &  1.41 &  
1.70 &  1.82 &  \phn1.51 &  \phn1.81 &  \phn1.91 &  \phn2.08 \\
E4 & \phn35--50\phn & 10.46 & 1.12 &  1.31 &   1.90 &  1.66 &  1.40 &  
1.59 &  1.53 &  \phn1.18 &  \phn1.62 &  \phn1.64 &  \phn2.15 \\
E5 & \phn50--75\phn & \phn9.75 & 1.02 &  1.30 &   1.88 &  1.27 &  1.59 &  
1.35 &  1.54 &  \phn1.28 &  \phn1.45 &  \phn1.41 &  \phn2.04 \\
E6 & \phn75--100 & 13.00 & 0.99 &  1.29 &   2.03 &  1.34 &  1.37 &  
1.52 &  1.25 &  \phn1.36 &  \phn1.51 &  \phn1.57 &  \phn1.85 \\
E7 & 100--150 & 24.48 & 1.75 &  1.31 &   2.67 &  1.40 &  1.47 &  1.59 &  
1.57 &  \phn1.31 &  \phn1.52 &  \phn1.48 &  \phn1.72 \\
E8 & 150--195 & 75.19 & 3.84 &  1.31 &   1.24 &  1.42 &  1.45 &  1.26 &  
1.29 &  \phn1.23 &  \phn1.65 &  \phn1.56 &  \phn1.93 \\ \hline
\multicolumn{4}{c}{Diffuse $N_{\rm PSF}$ (indep. of $E$):} & 1.00 & & &   
4.56 &   6.60 &   8.98 &  11.81 &  14.94 &  26.42 &  41.03
\enddata
\tablenotetext{a}{``Flux Calibration,'' defined such that the 
incidence photon flux at the Earth
(photons cm$^{-2}$ s$^{-1}$) is given by the BAT source count rate 
multiplied by Flux Calib.}
\tablenotetext{b}{Units are $10^{-5}$ photons s$^{-1}$ cm$^{-2}$ }
\end{deluxetable}

While this method is the standard way to create spectra from BAT survey
data, it is technically only valid for source spectra that have a shape
similar to the Crab.
Unfortunately, the BAT survey redistribution
matrix, which could properly account for arbitrary spectral shapes,
is more uncertain than the observed Crab fluxes \citep{Tue+10}.
Thermal emission above $\sim$10 keV is typically much steeper than the
spectral index in Equation~(\ref{eq:comabat:crab}), so we would prefer
to include an approximate redistribution matrix that will handle
other such spectral models correctly.
To do this, we take a standard response matrix for an
on-axis source from a single observation and multiply the input energies
by a smooth function so that the flux-converted Crab spectrum
matches Equation~(\ref{eq:comabat:crab}) to $<1$\% in all energy bands.
The addition of this redistribution matrix has a minor effect on
spectral fits generally, but it does improve the quality of fits
using a thermal model, so we employ it throughout.

\subsection{The Hard X-ray Spectrum of the Crab}
\label{sec:comabat:cn:hard}

With this approach, accurate conversions from BAT count rates to
true fluxes are not guaranteed.
The goal instead is to match the BAT calibration with the EPIC-pn
calibration, which ensures that spectral models can be applied seamlessly
between the \xmms and \swifts spectra in joint fits.
While fluxes quoted hereafter may differ from their true fluxes,
the relative amounts of thermal versus nonthermal emission 
in the joint spectra -- considering both their cross-normalization 
factor {\it and} shape -- are carefully conserved.
Ultimately, because our BAT calibration
method relies on using the Crab as a flux standard, and since the
true Crab spectrum is not known, the choice of a canonical
Crab spectrum is at some level arbitrary.

Even so, the hard band fluxes derived herein should be consistent with
fluxes derived from other missions.
Using the same power law form as for Equation~(\ref{eq:comabat:crab}),
\citet{Kir+05} found a range of normalizations and photon 
indices for several instruments that overlap with the BAT energy bands:
\saxs PDS: $8.84 E^{-2.126}$; \rxtes PCA: $11.02 E^{-2.120}$; \rxtes HEXTE:
$9.9 E^{-2.090}$; \integrals ISGRI: $15.47 E^{-2.252}$; and 
\integrals SPI: $15.9 E^{-2.203}$.
These photon indices agree well with the shape of the total Crab spectrum
determined by an earlier, detailed balloon-borne study, which found
$\Gamma=2.12$ \citep{Bar94}.
Also, the \suzakus PIN fit of $10.93 E^{-2.090}$ is consistent with both
the scatter in the above results and our adopted spectrum.
Ignoring the photon indices derived from \xmms data, there seems to be
a steepening in the Crab spectrum at higher energies, which means that
our relatively flat photon index may over-predict harder band fluxes and
thus enhance a potential nonthermal signal.
Our Crab spectrum also has a slightly higher overall flux (in the
20-80 keV band), so that fluxes and upper limits may be biased high,
though any such biasing would be well within the absolute calibration
uncertainties of all the above missions.

Assuming the \integrals SPI observations are the most accurate, due
to its extensive ground calibration, we can compare our adopted model
with the results of \citet{JR09}, which show a clear steepening of the
Crab spectrum above 100 keV.
Below 100 keV, they find a photon index of 2.07, consistent with our
value of 2.095.
Their best-fit photon index above 100 keV is 2.24, which would
lead our flux calibration to overestimate fluxes in our two highest energy
bands.
However, since we detect no flux in these bands, extended or otherwise,
even a large erroneous calibration -- which is not the case here --
would not bias or otherwise affect our results.
\citet{JR09} also fit a smoothly varying power law to their Crab spectrum.
Not surprisingly, this model is sufficiently similar to our adopted Crab
spectrum, particularly below 50 keV where the flux calibration is more
important,
that using it instead of our simple power law model would not lead to
quantitatively different results.

%------------------------------ DIFFUSE -------------------------------
\section{Extracting BAT Fluxes from Extended Sources} 
\label{sec:comabat:diffuse}

Very extended, diffuse emission is difficult to detect with
coded mask instruments, since the shadow pattern of the mask on the
detectors gets smeared out and the signal becomes indiscernible
from the background.
However, small scale extended emission is detectable, as long as its
size is less than the minimum scale necessary to dilute the
distinguishability of the mask pattern.
In the following, we show that, by simulating extended sources as
collections of point sources, this minimum scale is larger than
our region of interest and that essentially 100\% of the diffuse 
emission can be detected.

\subsection{BAT Point Spread Function}
\label{sec:comabat:diffuse:psf}

For on-axis sources in the BAT FOV, the point spread function (PSF) has
a full width half maximum (FWHM) of $\sim$22\arcmin.
Because the sky image is basically the cross-correlation function
of the coded mask with the count rates in the individual detectors, 
it does not represent the intensity per solid angle 
(i.e. within a pixel).
Instead, a pixel value is proportional to the flux of a point source
at that location.
The width of the PSF is actually due to
oversampling the sky plane, not the scattering of photons inside the
instrument, and it depends on the size of individual mask element
shadows on the detector relative to the size of detector pixels.

As such, the PSF should not be summed in order to derive the source
flux -- this is provided by the central peak value -- and its FWHM
depends on the off-axis angle of the source.
The distance between the mask and detector increases
as the off-axis angle increases,
so an angular separation at large off-axis angles produces a more
dramatic shift in the shadow pattern across the detector pixels than
more on-axis positions, which effectively reduces the oversampling
factor and leads to narrower FWHM.
Since survey images are created from many ``random'' individual pointings,
each with a given source located at a different off-axis angle,
the survey PSF will have an average FWHM and uniform shape, which is
roughly Gaussian.
Simple Gaussian fits to all the $\ga 10 \sigma$ sources in all 8 bands
yield an average FWHM of $19\farcm47$, irrespective of S/N or energy
band, essentially identical to the value determined in the 22-month survey
\citep{Tue+10}.

\begin{figure}
\plotone{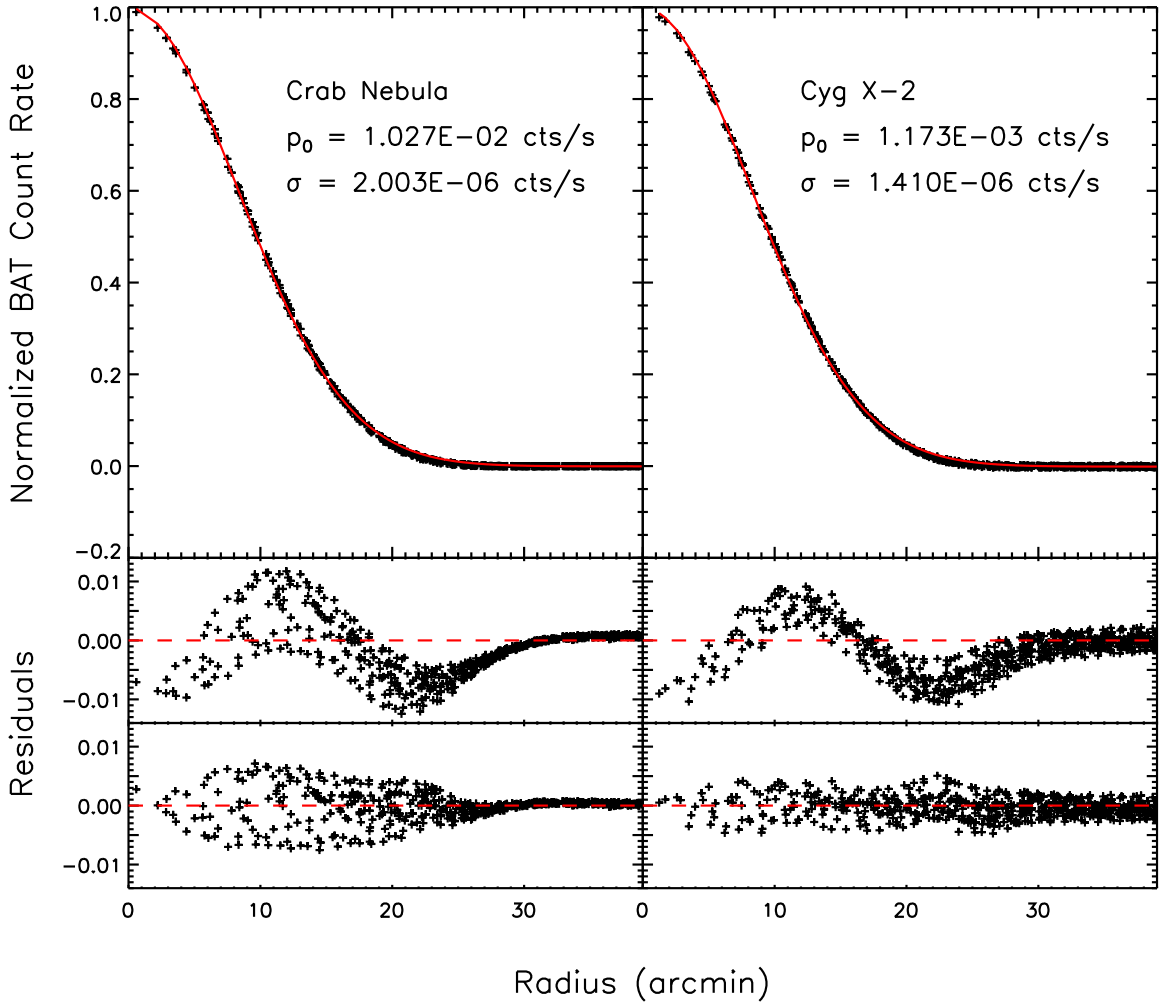}
\caption{Fits to the BAT PSF.  In the top and middle panels, the
profiles of two point sources, the Crab and Cyg X-2, are fit to a Gaussian.
Large, regular residuals (middle panels) remain, which are mostly
removed (at the $<1\%$ level, bottom panels) after modifying our 
expression for the PSF (Eqn.~\ref{eq:comabat:psf}).
Lingering residuals, which are particularly significant for the Crab
but only slightly noticeable for the other source, primarily result
from spatial asymmetries due to the actual angular extent of the
source, as in the case of the Crab, and the rectangular shape of
the BAT instrument, which will cause off-axis sources to be more
``squished'' in one direction than another.
In the latter case, the effect of summing many individual pointings
with the detector in various orientations
almost, but not entirely, removes this azimuthal component of the
PSF shape.
\label{fig:comabat:psffunc}}
\end{figure}

The PSF shape is described by a Gaussian to first order,
which is not surprising given the non-repeating, randomly filled mask
and the many pointings that contribute to the flux at each position.
However, as is clear from the residuals to a Gaussian fit to the 14--20 keV
band Crab data and another source (Cyg X-2)
in Figure~\ref{fig:comabat:psffunc}, 
deviations on the order of 1\% of the flux exist
(and are significantly larger than the root-mean square (RMS) of the 
background in this case).
While this deviation does not strongly impact the flux of point sources,
since only the maximum, central value maps to the flux,
a diffuse source is composed of overlapping PSFs, where differences
in the wings could affect the overall flux.
The residual structure in the wings of the PSF is mainly eliminated
by the addition of the two-part function:
\begin{equation} \label{eq:comabat:psf}
f(r) = p_0 \left[ e^{-r^2/2\sigma_{\rm PSF}^2} + \frac{1}{120}
\left\{ \begin{array}{ll}
\cos \pi x & x<3\\
-e^{-\pi(x-3)/1.19} & x>3
\end{array} \right. \right] + p_1
\, ,
\end{equation}
where $r$ is the distance from the center in arcminutes, $p_0$ and $p_1$
are fit parameters (the normalization and background, respectively), 
$x = 2 r / (1.19\, {\rm FWHM})+1$ ($x$ has units of radians), and
$\sigma_{\rm PSF} = {\rm FWHM} / (2\sqrt{2\ln2})$.
The improved fit for the Crab is illustrated in 
the very bottom left panel of Figure~\ref{fig:comabat:psffunc};
while in this case the fit is still not perfect, for other sources 
the fit is typically better (bottom right panel of
Figure~\ref{fig:comabat:psffunc}).
We take $p_0$ to be our measurement of the flux.
While the maximum of the PSF may not exactly correspond to $p_0$, since
all fluxes are determined this way and are also related to the Crab fluxes,
any such bias will cancel out during the conversion from BAT count rates
to fluxes.
Note that the most the additional terms to $f(r)$ could affect a flux,
assuming they, for some reason, poorly represented the true PSF shape,
is at the $\la 1$\% level.

% Figure 9

\subsection{Tests of the Detection of Extended Sources}
\label{sec:comabat:diffuse:tests}

As discussed in Section~\ref{sec:comabat:spatial},
we extract fluxes for extended sources by fitting  {\it a priori}
model distributions, as opposed to using a method like the ``CLEAN''
algorithm \citep{Hog74}, which reconstructs fluxes from an unknown 
underlining distribution assuming the PSF shape only.
The ``CLEAN'' method requires some fine-tuning, such as the region
of extraction (for clearly detected sources, expanding the 
region-of-interest even a little beyond the wings of the source can
significantly bias the derived flux), ``loop gain,'' and
completion threshold.
In our case, since there are only a few likely spatial distributions
for the thermal and any potential nonthermal emission, we are less
likely to produce biased fluxes by first assuming a spatial distribution
than by using a method like ``CLEAN.''
We represent a diffuse source as a collection of point sources, each
of which is convolved by the PSF (Eqn.~\ref{eq:comabat:psf}) and summed
together.

\begin{figure}
\plotone{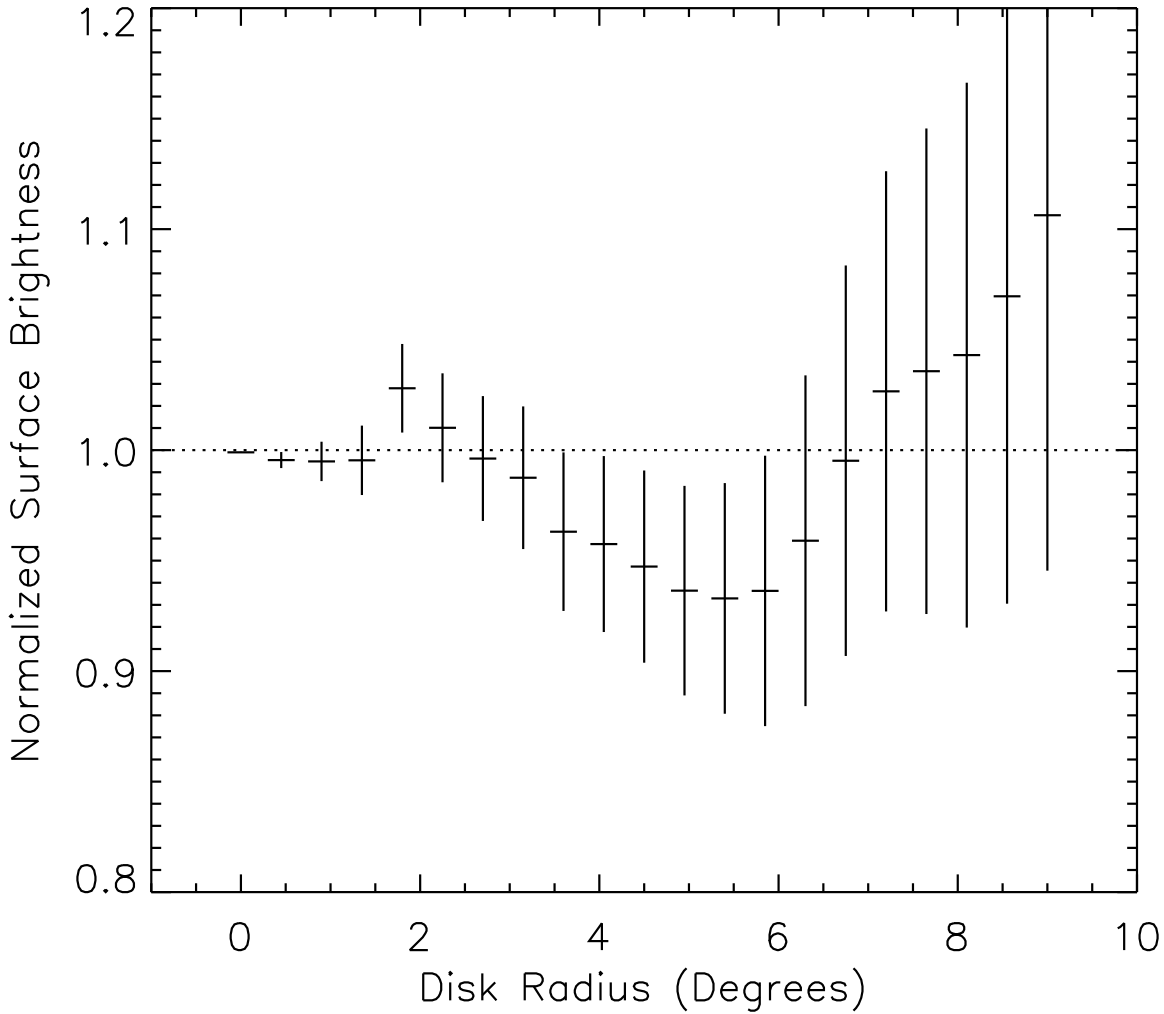}
\caption{Recovered surface brightness for simulated emission from
a uniform surface brightness disk of a given radius.
Error bars indicate the statistical error of the mean on the disk
surface brightness, though the spatial fluctuations in the sky
reconstruction behind this error are due entirely to systematic
effects; the ``noise'' at any position in these simulations is
determined by the flux of all the other sources within the FOV,
or in this case the other parts of the disk.
The variation with radius is smooth instead of random due to the
systematic origin of the fluctuations.
For angular sizes of interest here, $<90\arcmin$, the {\it intrinsic}
uncertainty in the recovered flux of an extended source, due to
coded mask imaging techniques, is at most a few percent
(based on the size of the error bars).
Also, there is {\it no} loss in sensitivity to diffuse emission;
{\it all} of the input flux is recovered, albeit with less and less
precision for larger sources.
\label{fig:comabat:diffsrc_recover}}
\end{figure}

We now test whether diffuse sources
are detectable over our scales of interest can be evaluated.
In general, we treat extended emission as a collection of closely-spaced
point sources, since existing software is built with these sources
in mind.
Point sources at any position in the BAT FOV are straightforward to 
simulate with the HEASOFT \swifts task {\tt batmaskwtimg} with the
following options set: {\tt coord\_type=tanxy}; {\tt distance=1e7}; 
{\tt corrections=forward,unbalanced,flatfield}; and {\tt rebalance=no}.
This task outputs the fraction of each detector pixel which is illuminated by
the source given its position relative to the detector axis; 
a value of 0.45 means that 55\% of
the detector area is shadowed by the mask.
At this stage, the detector image can be multiplied by the counts or
count rate of the source, and several such detector images representing 
different sources in the FOV can be added together along
with a background -- all including Poisson statistics.
The background can then be fit and subtracted with the task 
{\tt batclean}, and finally a reconstructed sky image can be
produced via the task {\tt batfftimage}.
For now, to isolate the detectability of diffuse emission by the BAT,
we simply add uniformly bright, perfectly known point source masks
without background or source Poisson noise,
to create circular, extended disks of various radii $R$.
Images of the sky are constructed with {\tt batfftimage} for each
disk detector image, and the ``observed'' disk surface brightness profile
is fit for as a function of radius.
While even large disks ($R \ga 10\arcdeg$) are visibly noticeable,
sky images produced in this way include significant systematic effects 
induced by the large spatial
extent of the emission, leading to high RMS noise that 
eventually destroys the disks' detectability.
The recovered surface brightness, relative to the input level, of simulated
diffuse disks of radius $R$ are presented
in Figure~\ref{fig:comabat:diffsrc_recover}.
Error bars represent the simple error of the mean 
($\sigma_{\rm RMS}/\sqrt{N}$, where $N$ is the number of pixels used to
determine $\sigma_{\rm RMS}$), and the smooth as opposed to random
variation around the input surface brightness results from their
systematic nature.
For sources in the size range of interest to us, $R<1\fdg5$, the
{\it intrinsic} uncertainty due to the telescope design is $\la 3\%$.
As the disk radius increases, the reconstructed surface brightness
becomes less and less robust as there are effectively more sources
(other parts of the disk) contributing systematic noise to a given
location.
Note, however, that {\it all} of the input source flux is recovered.

%Figure 10

%------------------------- FLUX UNCERTAINTIES ---------------------------
\section{Uncertainties in BAT Fluxes}
\label{sec:comabat:err}

\subsection{Flux Uncertainties for Point Sources}
\label{sec:comabat:err:pterr}

The uncertainty in a given flux measurement is encoded in the RMS
fluctuations in the local background \citep{Tue+10}.
These fluctuations represent both the statistical fluctuations from shot noise
(dominated by the high background rate) 
and systematic
error contributions from the sky reconstruction process.
Due to the large number of individual pointings at nearly 
random positions,
most systematic effects nearly average out and lead to a symmetric, 
nearly Gaussian
distribution for blank sky regions.
We calculate the RMS of the background ($\sigma_{\rm bgd}$) around 
Coma in an annulus of radius
$15<r<100$ pixels ($42\arcmin<r<4\fdg67$), as is typically done for 
sources in the BAT survey.
The values of $\sigma_{\rm bgd}$ for each band are given in 
Table~\ref{tab:comabat:err}.
While this annulus partially includes the region within which 
we are searching for a diffuse nonthermal signal, the lack of any
obvious emission indicates that the derived errors could not be
significantly biased.
To ensure $\sigma_{\rm bgd}$ is not biased by low level extended flux,
we recalculated it inside an annulus of equal area with an inner radius of
90\arcmin \, and found a nearly identical value of $\sigma_{\rm bgd}$
in all 8 bands.

\subsection{Flux Uncertainties for Extended Sources}
\label{sec:comabat:err:ext}

The error for a point source, or more correctly the error in the value
of a given pixel, presented in 
Section~\ref{sec:comabat:err:pterr} does not directly apply to
extended sources.
Also, we cannot take the standard error from spatial $\chi^2$ fits,
using the point source error as the error for the flux in each pixel,
because nearby pixels are correlated.
Helpfully, the expected error for diffuse sources has already been
derived by \citet{RGL+06} for the IBIS coded mask instrument onboard
\integral.
In their appendices, they derive source fluxes and errors in
reconstructed sky images from detector images and find the
straightforward result that the error in a flux measurement from
an extended
source is proportional to its spatial area normalized by the area
of the PSF function \citep[][Eqn.~B3]{RGL+06}.
Specifically, 
\begin{equation} \label{eq:comabat:differrarea}
\sigma_{\rm ext} = \sigma_{\rm bgd} \sqrt{N_{\rm PSF}}
\, ,
\end{equation}
where $N_{\rm PSF}$ is the area of the source normalized by the PSF area.
If $I_X ( \Omega )$ is the surface brightness of the source 
convolved with the PSF,
$I_X^{\rm max}$ is its maximum value,
$\Omega$ is the solid angle, and $f ( \Omega )$ is the PSF, then
\begin{equation} \label{eq:comabat:npsf}
N_{\rm PSF} \equiv \frac{ \int I_X ( \Omega ) \, d \Omega }
{ I_X^{\rm max} \int f ( \Omega ) \, d \Omega }
\, .
\end{equation}

\begin{figure}
\plotone{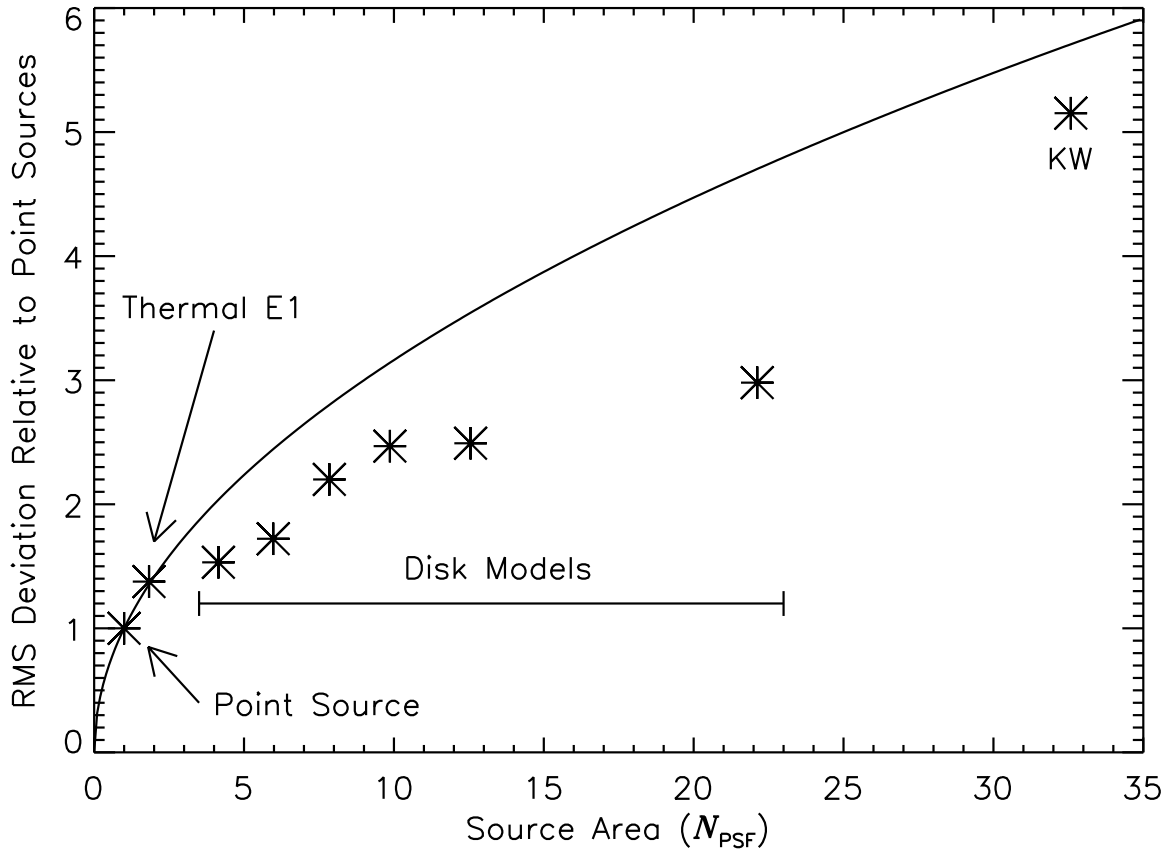}
\caption{The standard deviation of the best-fit normalization for
100 simulations of each spatial model considered in this work.
Both photon noise (in the background and source flux) and systematic
effects (the influence of other point sources in the FOV and the
relative off-axis angle of all sources relative to the detector)
are included.
The model area is shown in terms of the equivalent number of PSF
areas, $N_{\rm PSF}$.
For the thermal spatial models, only the lowest energy band 
(E1: 14--20 keV) is plotted for clarity.
The solid line represents the expected $\sqrt{N_{\rm PSF}}$ dependence
of the error 
(Equation~\ref{eq:comabat:differrarea}).
The simulated values of the errors for the extended models fall 
below this result, and possible explanations of
this behavior are briefly discussed in the text.
Note that for the thermal E1 model, the difference in the value of 
$N_{\rm PSF}$ shown here and in 
Table~\ref{tab:comabat:err} results from the difference between the
survey PSF and the on-axis PSF used in these simulations.
\label{fig:comabat:differr}}
\end{figure}

Before we generally apply Equation~(\ref{eq:comabat:differrarea}) to 
\swifts BAT data,
we test whether this prescription does in fact apply to extended 
sources in the \swifts BAT.
For each of the diffuse thermal 
(Section~\ref{sec:comabat:spatial:models:thermal}) or
nonthermal (Section~\ref{sec:comabat:spatial:models:nonthermal})
spatial models described above,
we created 121
simulations of the extended source and a number of other point
sources in the FOV, each with detector counts for the source fluxes
and background randomly assigned (taken from a Poisson distribution).
Each simulated observation is made from a unique position on an 
$11\times11$ grid, with the relative positions of all the sources kept
intact.
Both the variation of position relative to the telescope axis and
the inclusion of point sources are necessary to fully recreate the
systematic contribution to the error.
The simulated detector images are then background subtracted and
converted into sky images via the procedure outlined in 
Section~\ref{sec:comabat:diffuse:tests}.
The total flux of the diffuse sources is chosen so that the
signal-to-noise ratio is $\sim 20-50$.
To check Equation~(\ref{eq:comabat:differrarea}),
fluxes of all the sources are measured in each simulated sky image,
and the average standard deviation of the point source fluxes
are compared to the standard deviation of the diffuse source flux.
We find that the estimated errors for the
diffuse models (Thermal band E1, Disks R25--R60,
and KW) generally fall below the expected trend with $N_{\rm PSF}$
in Figure~\ref{fig:comabat:differr}.
This discrepancy may be due to the number of simulations
we were computationally limited to performing -- the distribution of
fluxes is only roughly Gaussian -- or it may represent a true
deviation from the results of \citet{RGL+06}.
However, to be safe we use Equation~(\ref{eq:comabat:differr}) to
calculate the error of fluxes extracted with the corresponding model.

%Figure 11

There is one additional modification to errors on fluxes extracted
with our methodology.
Because we fit a spatial model, convolved by the PSF, to the BAT
image data, the error in the flux is not just the standard deviation of
nearby background pixels, but it depends on how the model is fit to
all the pixels.
For example, the distribution of normalizations from many fits of
a Gaussian function
to random data of mean zero and standard deviation $\sigma_{\rm bgd}$ 
will not equal $\sigma_{\rm bgd}$ but some value $<\sigma_{\rm bgd}$ 
depending on the pixel scale.
A delta function, or Gaussian of width zero, will produce a
distribution consistent with $\sigma_{\rm bgd}$, since this is identical to
measuring the standard deviation, but anything wider
finds an average over several pixels, and therefore the distribution
of normalizations will tend to be closer to the mean of the random pixels.
For our purpose, where the normalization is related to the source flux,
the correct error of a flux should come from the distribution of model
fits to background (empty) regions of the survey, which may not be
equivalent to $\sigma_{\rm bgd}$.
Unlike in the above example, neighboring pixels in the survey are correlated 
due to oversampling -- this is essentially the origin of the PSF -- 
and so the standard deviation of model normalizations will be affected
by this correlation.
Generally, the distribution of normalizations will be larger than
$\sigma_{\rm bgd}$ in this case, as $\chi^2$ minimization will be more
influenced by the larger fluctuations near the flux extraction region.
The net effect does not significantly change the error distribution shape,
but simply inflates the effective standard deviation by some factor, $f_m$,
which is both model-dependent (varying from 1.4$\sigma_{\rm bgd}$ for a
point source to 2.24$\sigma_{\rm bgd}$ for the \kws model) and energy
dependent since the noise properties vary slightly from band-to-band.
The total flux uncertainty for a diffuse source in the BAT survey is
adjusted from Equation~(\ref{eq:comabat:differrarea}) to become
\begin{equation} \label{eq:comabat:differr}
\sigma_{\rm diffuse} = f_m \sigma_{\rm ext} = f_m \sigma_{\rm bgd} 
\sqrt{N_{\rm PSF}}
\, .
\end{equation}
The precise value of $f_m$ is determined from the standard deviation of
fits to 100 blank sky regions, in which we avoid obvious ($>5\sigma$)
sources and the Galactic plane ($b>20\arcdeg$).
These factors are reported for each band in Table~\ref{tab:comabat:err}.
Not including this factor in the flux error estimate 
results in spectral fits with
unacceptably high $\chi^2$ values.

%-------------------------- REFERENCES --------------------------------


\begin{thebibliography}{46}
\expandafter\ifx\csname natexlab\endcsname\relax\def\natexlab#1{#1}\fi

\bibitem[Ajello et al.(2009)]{Aje+09} Ajello, M., et al.\ 
2009, \apj, 690, 367

\bibitem[Arnaud et al.(2001)]{Arn+01} Arnaud, M., et al.\ 2001, \aap,
365, L67

\bibitem[Bartlett(1994)]{Bar94} Bartlett, L.~M.\ 1994, 
Ph.D.~Thesis

\bibitem[Bazzano et al.(1990)]{Baz+90} Bazzano, A., et al.\
1990, \apjl, 362, L51

\bibitem[Briel et al.(2001)]{Bri+01} Briel, U.~G., et al.\ 2001, \aap,
365, L60

\bibitem[Bonafede et al.(2010)]{BFM+10} Bonafede, A., Feretti, L., 
Murgia, M., Govoni, F., Giovannini, G., Dallacasa, D., Dolag, K., 
\& Taylor, G.~B.\ 2010, \aap, 513, A30

\bibitem[Brunetti \& Lazarian(2010)]{BL10} 
Brunetti, G., \& Lazarian, A.\ 2010, \mnras, 1371

\bibitem[Cavagnolo et al.(2008)]{CDV+08} Cavagnolo, K.~W., 
Donahue, M., Voit, G.~M., \& Sun, M.\ 2008, \apj, 682, 821

\bibitem[Deiss et al.(1997)]{DRL+97} Deiss, B.~M., Reich, W., Lesch, H.,
\& Wielebinski, R.\ 1997, \aap, 321, 55

\bibitem[Dolag et al.(2008)]{DBD08} Dolag, K., Bykov, A.~M., 
\& Diaferio, A.\ 2008, \ssr, 134, 311

\bibitem[Eckert et al.(2007)]{ENC+07} Eckert, D., Neronov, A., Courvoisier,
T.~J.-L., \& Produit, N.\ 2007, \aap, 470, 835

\bibitem[Eckert et al.(2008)]{EPP+08} Eckert, D., Produit, N., Paltani, S.,
Neronov, A., \& Courvoisier, T.~J.-L.\ 2008, \aap, 479, 27

\bibitem[Feretti et al.(1995)]{FDG+95} Feretti, L., Dallacasa, D.,
Giovannini, G., \& Tagliani, A.\ 1995, \aap, 302, 680

\bibitem[Fujita et al.(2008)]{Fuj+08} Fujita, Y., et al.\ 
2008, \pasj, 60, 1133

\bibitem[Fusco-Femiano et al.(1999)]{FDF+99} Fusco-Femiano,
R., dal Fiume, D., Feretti, L., Giovannini, G., Grandi, P., Matt, G.,
Molendi, S., \& Santangelo, A.\ 1999, \apjl, 513, L21

\bibitem[Fusco-Femiano et al.(2004)]{FOB+04} Fusco-Femiano, 
R., Orlandini, M., Brunetti, G., Feretti, L., Giovannini, G., Grandi, P., 
\& Setti, G.\ 2004, \apjl, 602, L73

\bibitem[Fusco-Femiano et al.(2007)]{FLO07} Fusco-Femiano, 
R., Landi, R., \& Orlandini, M.\ 2007, \apjl, 654, L9

\bibitem[Giovannini et al.(1993)]{GFV+93} Giovannini, G.,
Feretti, L., Venturi, T., Kim, K.-T.,
\& Kronberg, P.~P.\ 1993, \apj, 406, 399

\bibitem[Harris \& Romanishin(1974)]{HR74} Harris, D.~E., \& Romanishin, W.\ 
1974, \apj, 188, 209

\bibitem[Henriksen \& Mushotzky(1986)]{HM86} Henriksen, M.~J.,
\& Mushotzky, R.~F.\ 1986, \apj, 302, 287

\bibitem[H{\"o}gbom(1974)]{Hog74} H{\"o}gbom, J.~A.\ 1974, \aaps, 15, 417

\bibitem[Hughes et al.(1993)]{HBS+93} Hughes, J.~P., Butcher, 
J.~A., Stewart, G.~C., \& Tanaka, Y.\ 1993, \apj, 404, 611 

\bibitem[Jourdain \& Roques(2009)]{JR09} 
Jourdain, E., \& Roques, J.~P.\ 2009, \apj, 704, 17

\bibitem[Kirsch et al.(2005)]{Kir+05} Kirsch, M.~G., et al.\ 
2005, \procspie, 5898, 22

\bibitem[Kushnir \& Waxman(2010)]{KW10} Kushnir, D.,
\& Waxman, E.\ 2010, Journal of Cosmology and Astro-Particle 
Physics, 2, 25

\bibitem[Lutovinov et al.(2008)]{LVC+08} Lutovinov, A.~A., 
Vikhlinin, A., Churazov, E.~M., Revnivtsev, M.~G., 
\& Sunyaev, R.~A.\ 2008, \apj, 687, 968

\bibitem[Madsen et al.(2009)]{MHK+09} Madsen, K., Harrison, 
F., Koglin, J., Mao, P., Craig, W., Pivovaroff, M., 
\& Christensen, F.\ 2009, Bulletin of the American Astronomical Society, 
41, 347

\bibitem[Mantz et al.(2008)]{MAE+08} Mantz, A., Allen, S.~W., 
Ebeling, H., \& Rapetti, D.\ 2008, \mnras, 387, 1179

\bibitem[Markwardt(2009)]{Mar09} Markwardt, C.~B.\ 2009, 
Astronomical Society of the Pacific Conference Series, 411, 251

\bibitem[Million \& Allen(2009)]{MA09} Million, E.~T., \& Allen, 
S.~W.\ 2009, \mnras, 399, 1307

\bibitem[Murgia et al.(2004)]{Mur+04} Murgia, M., Govoni, F., 
Feretti, L., Giovannini, G., Dallacasa, D., Fanti, R., Taylor, G.~B., 
\& Dolag, K.\ 2004, \aap, 424, 429

\bibitem[Murgia et al.(2010)]{Mur+10} Murgia, M., Eckert, D., 
Govoni, F., Ferrari, C., Pandey-Pommier, M., Nevalainen, J., \& 
Paltani, S.\ 2010, \aap, 514, A76

\bibitem[Nevalainen et al.(2004)]{NOB+04} Nevalainen, J.,
Oosterbroek, T., Bonamente, M., \& Colafrancesco, S.\ 2004, \apj, 608, 166

\bibitem[Nevalainen et al.(2010)]{NDG10} Nevalainen, J., 
David, L., \& Guainazzi, M.\ 2010, arXiv:1008.2102

\bibitem[Petrosian(2001)]{Pet01} Petrosian, V.\ 2001, \apj, 557, 560

\bibitem[Read \& Ponman(2003)]{RP03} Read, A.~M., \& Ponman, T.~J.\ 
2003, \aap, 409, 395

\bibitem[Rephaeli(1977)]{Rep77} Rephaeli, Y.\ 1977, \apj,
212, 608

\bibitem[Renaud et al.(2006)]{RBP+06} Renaud, M., B{\'e}langer, G., 
Paul, J., Lebrun, F., \& Terrier, R.\ 2006, \aap, 453, L5

\bibitem[Renaud et al.(2006)]{RGL+06} Renaud, M., Gros, A., Lebrun, F., 
Terrier, R., Goldwurm, A., Reynolds, S., \& Kalemci, E.\ 
2006, \aap, 456, 389

\bibitem[Rephaeli et al.(1994)]{RUG94} Rephaeli, Y., Ulmer,
M., \& Gruber, D.\ 1994, \apj, 429, 554

\bibitem[Rephaeli \& Gruber(2002)]{RG02} Rephaeli, Y., \& Gruber, D.\ 2002,
\apj, 579, 587

\bibitem[Rossetti \& Molendi(2004)]{RM04} Rossetti, M., \& Molendi, S.\
2004, \aap, 414, L41

\bibitem[Sarazin(1988)]{Sar88} Sarazin, C.~L.\ 1988, 
Cambridge Astrophysics Series, Cambridge: Cambridge University Press, 1988

\bibitem[Schuecker et al.(2004)]{SFM+04} Schuecker, P., Finoguenov, A.,
Miniati, F., B{\"o}hringer, H., \& Briel, U.~G.\ 2004, \aap, 426, 387

\bibitem[Thierbach et al.(2003)]{TKW03} Thierbach, M., Klein, U., \&
Wielebinski, R.\ 2003, \aap, 397, 53

\bibitem[Tueller et al.(2010)]{Tue+10} Tueller, J., et al.\ 
2010, \apjs, 186, 378

\bibitem[Vazza et al.(2009)]{VBG09} Vazza, F., Brunetti, G., 
\& Gheller, C.\ 2009, \mnras, 395, 1333

\bibitem[Vanderlinde et al.(2010)]{Van+10} Vanderlinde, K., et 
al.\ 2010, arXiv:1003.0003

\bibitem[Vikhlinin et al.(2009)]{Vik+09} Vikhlinin, A., et 
al.\ 2009, \apj, 692, 1060

\bibitem[Weisskopf et al.(2010)]{WGJ+10} Weisskopf, M.~C., 
Guainazzi, M., Jahoda, K., Shaposhnikov, N., O'Dell, S.~L., Zavlin, V.~E., 
Wilson-Hodge, C., \& Elsner, R.~F.\ 2010, \apj, 713, 912

\bibitem[Wik et al.(2009)]{WSF+09} Wik, D.~R., Sarazin, C.~L., 
Finoguenov, A., Matsushita, K., Nakazawa, K., 
\& Clarke, T.~E.\ 2009, \apj, 696, 1700

\bibitem[Willson(1970)]{Wil70} Willson, M.~A.~G.\ 1970,
\mnras, 151, 1

\end{thebibliography}
\end{document}